\documentclass[prb,aps,amsmath,amssymb,superscriptaddress,twocolumn,longbibliography]{revtex4-2}
\usepackage{graphicx,multirow}
\usepackage{subcaption}
\usepackage{tikz}
\usepackage{amsmath}
\usepackage{color,outlines}
\usepackage[english]{babel}
\usepackage{amsthm}
\usepackage{hyperref}
\usepackage{physics}
\usepackage{outlines}
\usepackage{float}
\usepackage[normalem]{ulem}

\definecolor{green2}{RGB}{0,100,0}

\begin{document}

\title{Emergence of Hermitian topology from non-Hermitian knots}
\author{Gaurav Hajong}
\email{gauravhajong730@gmail.com}
\affiliation{Department of Physics, Banaras Hindu University, Varanasi~221005, India}
\author{Ranjan Modak}
\email{ranjan@iittp.ac.in}
\affiliation{Indian Institute of Technology Tirupati, Tirupati, India~517619}  
\author{Bhabani Prasad Mandal}
\email{bhabani@bhu.ac.in }
\affiliation{Department of Physics, Banaras Hindu University, Varanasi~221005, India}

\begin{abstract}

The non-Hermiticity of the system gives rise to a distinct knot topology in the complex eigenvalue spectrum, which has no counterpart in Hermitian systems. In contrast, the singular values of a non-Hermitian (NH) Hamiltonian are always real by definition, meaning that they can also be interpreted as the eigenvalues of some underlying Hermitian Hamiltonian. In this work, we demonstrate that if the singular values of an NH Hamiltonian are treated as eigenvalues of prototype translational invariant Hermitian models that undergo a topological phase transition between two distinct topological phases, the complex eigenvalues of the NH Hamiltonian will also undergo a {\it{first order knot transition}} between different knot structures. Unlike the usual knot transition,  this transition is not accompanied by an Exceptional point (EP); in contrast, the real and complex parts of the eigenvalues of the NH Hamiltonian show a discrete jump at the transition point. 
We emphasize that the choice of an NH Hamiltonian whose singular values match the eigenvalues of a Hermitian model is not unique. However, our study suggests that this connection between the NH and Hermitian models remains robust as long as the periodicity in lattice momentum is the same for both. Furthermore, we provide an example showing that a change in the topology of the Hermitian model implies a transition in the underlying NH knot topology, but a change in knot topology does not necessarily signal a topological transition in the Hermitian system.
\end{abstract}

\maketitle

\section{Introduction} 
A consistent quantum theory with a complete real spectrum, unitary time evolution, and probabilistic interpretation for a certain class of NH systems~\cite{PhysRevLett.80.5243,Bender_2002,pub.1016488528,doi:10.1142/S0219887810004816} has been developed in a modified Hilbert space equipped with an appropriate inner product~\cite{mostafazadeh2003pseudo,AMD2010pseudo,PhysRevA.100.062118,PhysRevResearch.3.013015}. This brought much attention to understanding NH systems in the early 21st century. In the last few years, the interest in NH systems has grown exponentially due to their extensive application in PT phase transitions~ \cite{KHARE200053,MANDAL2015185,RAVAL2019114699,MANDAL20131043,PhysRevE.111.014421}, photonics~\cite{peng2014parity,PhysRevLett.113.053604,PhysRevLett.115.040402,PhysRevLett.100.103904,regensburger2012parity,feng2017non,feng2014single}, condensed-matter systems~\cite{PhysRevLett.121.026403,PhysRevB.98.035141,PhysRevB.99.201107,PhysRevA.110.042607}, and open quantum systems~\cite{PhysRevLett.70.2273,zhen2015spawning,diehl2011topology}, and various interdisciplinary areas~\cite{PhysRevA.109.022227,PhysRevA.103.062416,PhysRevE.111.014421,un_23,modak2023non, BASUMALLICK2001231,Hasan_2018,hasan2020, Ghatak_2015}. NH Hamiltonians have complex eigenenergies and that 
give rise to a number of interesting phenomena that Hermitian systems do not have, for example, the presence of an EP, where eigenvalues along with eigenstates coalesce~\cite{miri2019exceptional,heiss2012physics},
and the NH skin effect, where an extensive number of eigenmodes are localized at the boundary~\cite{PhysRevLett.124.086801,PhysRevLett.124.056802}.  Understanding the topology introduced by the NH Hamiltonian was initially centered around the NH band theory~\cite{RevModPhys.93.015005,PhysRevLett.123.066404}, and the classification of topologically
distinct NH Hamiltonians was done based on symmetry, akin
to the Hermitian tenfold way~\cite{RevModPhys.88.035005}. This helps one to explore topologically
robust quantities, nontrivial edge states, and many intriguing phenomena like NH bulk boundary correspondence \cite{PhysRevB.103.075126,NNHB,10.21468/SciPostPhys.15.4.173}, 
that also explains the NH skin effect~\cite{PhysRevLett.125.126402,ntrejk}.

Recently, it was recognized that the NH band theory 
is incomplete~\cite{PhysRevX.8.031079,PhysRevB.99.235112,PhysRevX.9.041015}, 
leading to the introduction of a framework based on homotopy theory to classify the NH
topological phases~\cite{PhysRevB.101.205417,PhysRevLett.126.010401,PhysRevB.103.075441,PhysRevB.105.195407,PhysRevB.109.094311,PhysRevLett.127.213601,tyjp}. This classification method does not rely on specific symmetries; in contrast, 
it is done by a mapping from the Brillouin zone to the space of energy bands and eigenstates.
The non-Hermiticity of the Hamiltonian can generate complex eigenvalues, which can constitute the knot structures \cite{PhysRevLett.134.126603,NC,knottt,tssafo,PhysRevLett.130.017201,PhysRevLett.124.186402}. The classification of NH systems with knot topology
is done based on the braiding of
eigenvalues, and further classified based on the
eigenstate topology~\cite{PhysRevB.103.155129}. Several experimental studies have recently aimed at revealing the knot topology of the NH systems; for example, the braiding structure of energy bands has been observed in
optical, mechanical, and trapped ion systems~\cite{PhysRevLett.130.163001,wang2021topological,patil2022measuring}. 

Our work focuses on singular values, instead of complex eigenvalues, in  NH Hamiltonians. For a given matrix $A$, its singular values are defined as the square roots of the eigenvalues of $AA^{\dag}$. If  $A$ is Hermitian, singular values reduce to the absolute values of the eigenvalues and hence contain essentially the same information as eigenvalues. 
In contrast, for NH matrices $A$, singular values play a distinct role compared to eigenvalues, and these have broad applications in diverse areas of science, such as in various tensor-network techniques \cite{article,PhysRevResearch.7.013237,PhysRevB.97.125102,PhysRevE.97.033310,campbell2024bridgingrelationssvdtensor}, CP violation \cite{Cheng:1984vwu}, Renormalisation Group analysis \cite{PhysRevC.104.044002}, reflection, and transmission probability~\cite{datta1997electronic}, chaos theory and applications \cite{PhysRevB.111.L161408,baggioli2025singularvaluedecompositionblind,PhysRevB.111.L060201,PhysRevB.111.064203}. While complex eigenvalues and consequently, an NH Hamiltonian, are crucial for the emergence of knot structures in the eigenvalue spectrum, one might wonder: Can the singular values of NH matrices have any imprint of such knot structures on the underlying complex eigenvalue spectrum? It's important to note that singular values are always real and thus cannot exhibit knot topology. However, they can be interpreted as the eigenvalues of a Hermitian Hamiltonian and can shed some light on the topological behavior of the Hermitian system. In this manuscript, we pose a related question: If a Hermitian Hamiltonian undergoes a topological phase transition between distinct phases by varying a tunable parameter, and we assume that the eigenvalues of the Hermitian Hamiltonian are the singular values of an underlying NH Hamiltonian, can tuning the parameter across different topological phases in the Hermitian system induce a change in knot topology in the corresponding NH Hamiltonian? We answer the question affirmatively here; we choose a prototype Hermitian extended Su–Schrieffer–Heeger (SSH) model that goes through different topological phase transitions~\cite{ssh}, and construct different underlying NH Hamiltonians whose singular values are the same as the eigenvalues of the Hermitian model. It is to be noted that there have been earlier works which have explored connections between Hermitian and NH topologies~\cite{PhysRevLett.133.266604,PhysRevLett.123.206404,PhysRevB.108.165105}.
It was further demonstrated that the braiding and linking of complex-energy bands of a given NH operator can encode the winding topology of a chiral Hermitian Hamiltonian constructed from it, thereby providing a braiding-based perspective on Hermitian invariants~\cite{PhysRevB.108.165105}. In contrast, the present work focuses on a \emph{family} of NH matrices \(A\) generated via the SVD construction, which naturally incorporates gauge freedom, rather than on a fixed NH operator. We also note that there have been earlier works which have used SVD-based classifications for NH systems in different context~\cite{PhysRevA.99.052118}. Our main finding is that the complex eigenvalues of the NH Hamiltonians also go through a change in knot topology precisely at the same point where the Hermitian Hamiltonians display a topological transition. However, this transition is not accompanied by an EP or non-defective degeneracy point (NDP)~\cite{PhysRevLett.126.010401}; in contrast, the real and complex parts of the eigenvalues of the NH Hamiltonian show a discrete jump at the transition point,
which we refer to as {\it{first order knot transition}}.

\section{Formalism}\label{Formalism}
We consider a translational invariant $n\times n$ Hermitian Hamiltonian $H(k,\omega)$ (where $k\in [0, 2\pi]$ is lattice momentum)  such that its energy eigenvalues $\mathcal{E}_i>0$, $i=1,2,3,\cdots n $ and also, by tuning the parameter $\omega$, the Hamiltonian goes through topological phase transition between distinct topological phases. 
Strictly speaking, to ensure that the eigenvalues of $H$ are real and positive, we add a term $\mu I$ to the Hamiltonian, where $I$ is the $n\times n$ identity matrix and $\mu$ is an appropriately chosen constant that guarantees the positivity of the spectrum. Accordingly, we write
\[
H = H_{\mathrm{TP}} + \mu I,
\]
where $H_{\mathrm{TP}}$ possesses chiral symmetry and undergoes a change in the winding number in the usual sense, thereby exhibiting a topological transition. Throughout this manuscript, whenever we refer to a topological transition in $H$, we strictly mean the transition of $H_{\mathrm{TP}}$.

Given $H=H^{\dag}$, it can be always expressed as matrix product of two $n\times n$ matrices, i.e., $A$ and its Hermitian conjugate $A^\dagger$,
\begin{equation}\label{HTN}
    H(k,\omega)=A(\omega,k){A^\dagger(\omega,k)}, 
    \end{equation}
where $A$ can even be NH. It is straightforward to see that $A$ can be constructed using the method of singular value decomposition (SVD) as \cite{Cheng:1984vwu}, \begin{equation}\label{SVD}
    A(\omega,k)= U\Sigma{{V}^\dagger}, 
    \end{equation}
where $U$ is a unitary matrix constructed out of the eigenvectors of the parent Hermitian Hamiltonian $H$ and $\Sigma=\text{diag}(\sqrt{\mathcal{E}_1}, \sqrt{\mathcal{E}_2},\cdots \sqrt{\mathcal{E}_n})$ is a diagonal matrix whose elements are the square roots of the eigenvalues of $H$. $V$ is also unitary but arbitrary matrix. Hence, $A$ is not unique for a given $H$ and one can construct infinitely many $A$'s by choosing different $V$. Additionally, we also note that with the given condition of Eqs.~\eqref{HTN} and \eqref{SVD}, we can generalize it to include more general $A'$ as well through the relation, $A'=A\mathcal{U}$, such that $\mathcal{U}$ is a general unitary matrix.

To distinguish different knot-structures for the NH Hamiltonian $A$, we use the following definition of the winding number \cite{PhysRevX.8.031079,wang2021topological}, 
\begin{equation}\label{winding-number}
    \nu=\int^{2\pi}_{0}\frac{dk}{2\pi{i}}\frac{d}{dk}\ln \text{Det}\big\{A-\frac{1}{2}\Tr A\big\}. 
\end{equation}
Note that the term $\tfrac{1}{2}\Tr A$ in Eq.~\eqref{winding-number} provides the appropriate reference energy for computing the winding number in a two-band ($2\times2$) system, which is the case for all models studied in this work. For a general $n\times n$ NH matrix, the natural extension is to use $\tfrac{1}{n}\Tr A$, ensuring that the winding number is defined with respect to the spectral center. Different knot structures in the complex eigenvalue spectrum correspond to different $\nu$, such as for the unlink phase $\nu=0$, the unknot phase corresponds to $\nu=1$, and   $\nu=2$ for the Hopf-link phase~\cite{wu2023observation}. While on the other hand, for the Hermitian Hamiltonian $H(\omega,k)$, the topological invariant is the standard one-dimensional winding number defined from the eigenvectors. Denoting the eigenvector by $|u_\pm(k)\rangle$, we have now
\begin{equation}\label{nu-H}
\nu_{\mathrm H}=\frac{i}{2\pi}\int_{0}^{2\pi}\langle u_\pm(k)|\partial_k u_\pm(k)\rangle\,dk .
\end{equation}
This definition is unaffected by identity shifts such as the $\mu$ term, since they do not modify the eigenvectors.

Typically, the knot transition is always accompanied by an EP. On the other hand, energy gap-closing is essential for the Hermitian $H$ at the topological transition point. Gap-closing implies the degeneracies in $H$. 
Then, at least for the two-bands model (which we are going to focus on in this manuscript), $A= \epsilon ^{1/2}U{{V}^\dagger}$ (replacing $\mathcal{E}_1=\mathcal{E}_2=\epsilon$ in Eq.~\eqref{SVD} and $n=2$). It automatically implies, $A^{\dag}= \epsilon ^{1/2}V{{U}^\dagger}$ and $[A,A^{\dag}]=0$.  Consequently, $A$ is a normal matrix, diagonalizable; EP does not exist. The absence of an EP at the transition point would typically suggest that a knot transition is unlikely. However, our results would indicate otherwise.

\section{Models}\label{Models}
The Hermitian Hamiltonian we consider here is the one-dimensional SSH model and an extended version of it.  The SSH system is one of the simplest topological
systems~\cite {ssh}, which plays a crucial role in understanding topological transitions in Hermitian systems, and there have also been several interesting
works on its experimental realization, which makes this model even more compelling~\cite{RevModPhys.91.015005,kiczynski2022engineering,meier2016observation,PhysRevB.105.195419}. The Hamiltonian is given by~\cite{PhysRevB.110.054312}, 
\begin{eqnarray}
      H=\sum_{i=1}^{L}( t_1 c^{\dag}_{i,A}c_{i,B}+t_2 c^{\dag}_{i,B}c_{i+1,A}+t_3 c^{\dag}_{i,A}c_{i+1,B}\nonumber \\ 
    +t_4 c^{\dag}_{i,B}c_{i+2,A} +\text{H.c.})+\mu\sum_{i=1}^{L}(n_{i,A}+n_{i,B}),
    \label{model}
\end{eqnarray}
 where $L$ is the number of unit cells, A and B denote the two
sub-lattices, and  $c^{\dag}_{i, X}$ ($c_{i, X}$) corresponds to a fermionic creation (annihilation) operator, and  $n_{i, X}=c^{\dag}_{i,X} c_{i, X}$ is occupation operator on $i$-th unit cell in the sub-lattice X. 
$t_1$ and $t_2$  are the intracell and intercell hoppings in the original SSH model, and the $t_3$ and $t_4$ terms are added to include chiral-symmetry-preserving next-nearest-neighbor hopping. Given a choice of parameters of hopping strengths, $\mu$ will be chosen in such a way that the eigenvalues are always real and positive, i.e., $\geq0$. As mentioned above,
we are interested in a Hermitian Hamiltonian such that its spectrum can be thought of as singular values of some NH Hamiltonian, and singular values are by construction always $\geq 0$, which enforces us to make this choice, however, it does not change physics of topological phases in the SSH model, given this term corresponds to just a constant shift in energy. 

The model can be solved exactly via the Fourier transformation as $c_{k,A} = \frac{1}{\sqrt{L}} \sum_j e^{-ikj} c_{j,A}$, and $c_{k,B} = \frac{1}{\sqrt{L}} \sum_j e^{-ikj} c_{j,B}$,
which brings the Hamiltonian of the system into the form of $  H = \sum_k \Psi_k^\dagger H(k) \Psi_k$,
where $\Psi_k = ( c_{k, A},c_{k, B} )^T$ and $H(k)= \vec{d}(k) \cdot \vec{\sigma}+\mu \mathbb{I}$ is the Bloch Hamiltonian  where the associated Bloch vector has the components
\begin{equation}
    \begin{aligned}
        d_x(k) &= t_1 + ( t_2 + t_3 )\cos k + t_4\cos( 2k );\\
        d_y(k) &= ( t_2 - t_3 )\sin k + t_4\sin( 2k );\ 
        d_z(k) = 0
    \end{aligned}
\end{equation}
and $\vec{\sigma}$ is the Pauli matrix vector. 
In this manuscript, we consider two scenarios: (1) Model I:  $t_1=1$, $t_2=\omega$, $t_3=t_4=0$, and $\mu=1+\omega$, (2) Model II: $t_1=1$, $t_2=1$, $t_3=1$, $t_4=\omega$, and $\mu=3+\omega$. Using Eq.~\eqref{winding-number}, it is straightforward to see that the model I shows a topological transition between two topologically distinct phases, $\nu=0$ phase to $\nu=1$ phase at $\omega=1$, and on the other hand, the model II shows a topological transition between $\nu=1$ phase to $\nu=2$ phase at $\omega=1$. 

\section{Results}\label{results}

We now demonstrate our results for both the models.
Note that the Hermitian Hamiltonians we consider for both the models are periodic in lattice momentum $k$, i.e., $H(k+2\pi)=H(k)$. Hence we restrict ourselves to $A$ with same periodic behavior, $A(k+2\pi)=A(k)$. Among the infinitely many $A$'s as indicated by Eq.~\eqref{SVD} for arbitrary $V$, we first consider those which are constructed using traceless, $k$-independent $V$ for simplicity. The consequence of more complicated cases, including $k$-dependent choices of $V$ and the resulting broader family of gauge-related matrices $A'$, are demonstrated in {\bf Appendix~\eqref{appa}}.


\subsection{Model I}\label{MM-1}
\begin{figure}
  \begin{subfigure}{0.48\columnwidth}
  \includegraphics[width=\textwidth]{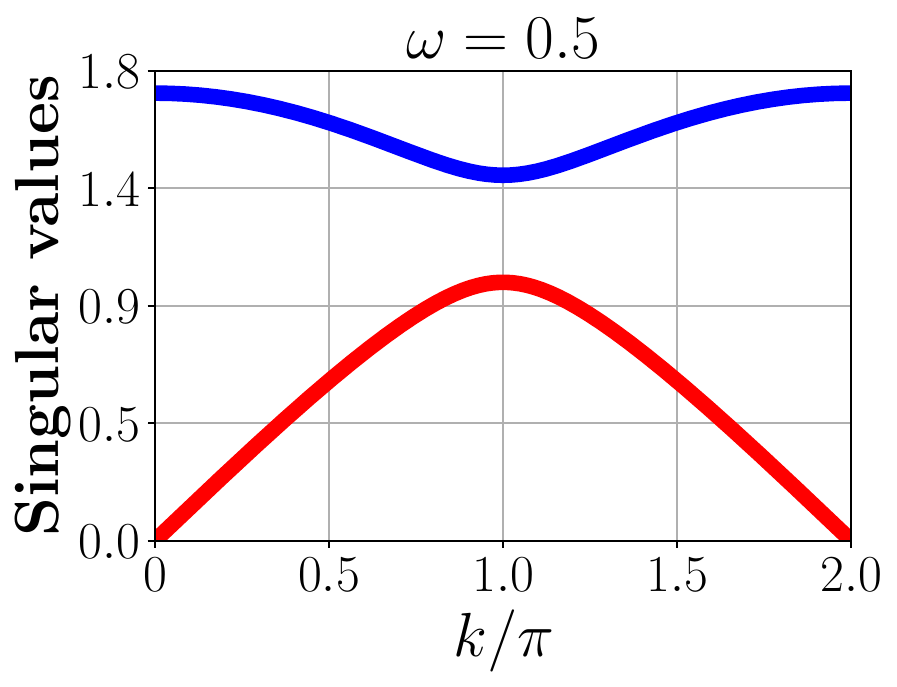}
  \caption{}\label{HA-0_5}
  \end{subfigure}
  \begin{subfigure}{0.48\columnwidth} 
  \includegraphics[width=\textwidth]{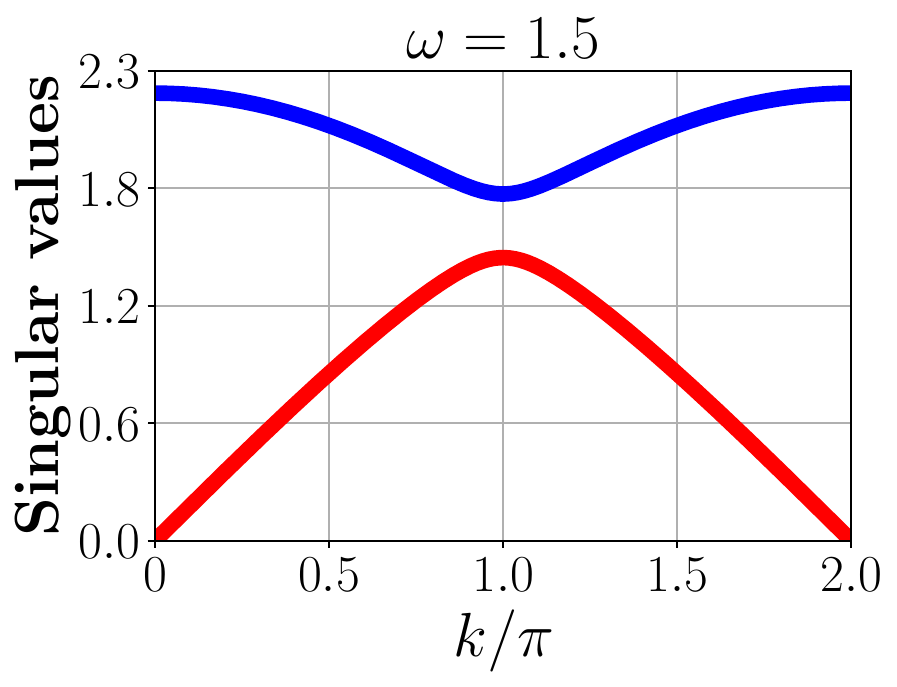} 
  \caption{}\label{HA-1_5} 
  \end{subfigure}
  \vspace{0pt}
  \begin{subfigure}{0.48\columnwidth} 
  \includegraphics[width=\textwidth]{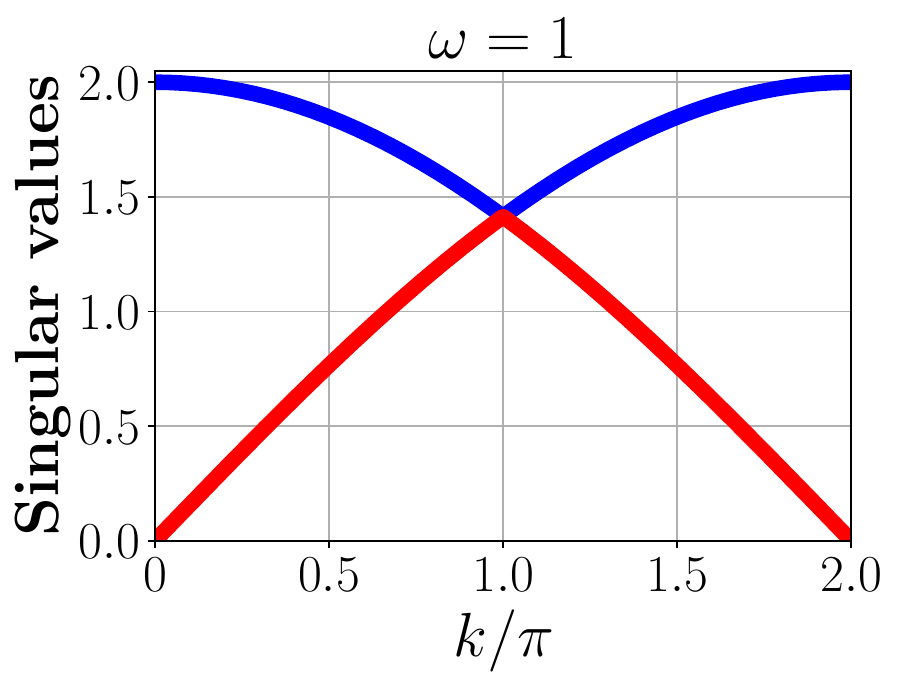} 
  \caption{}\label{HA-1} 
  \end{subfigure}
  \begin{subfigure}{0.48\columnwidth} 
  \includegraphics[width=\textwidth]{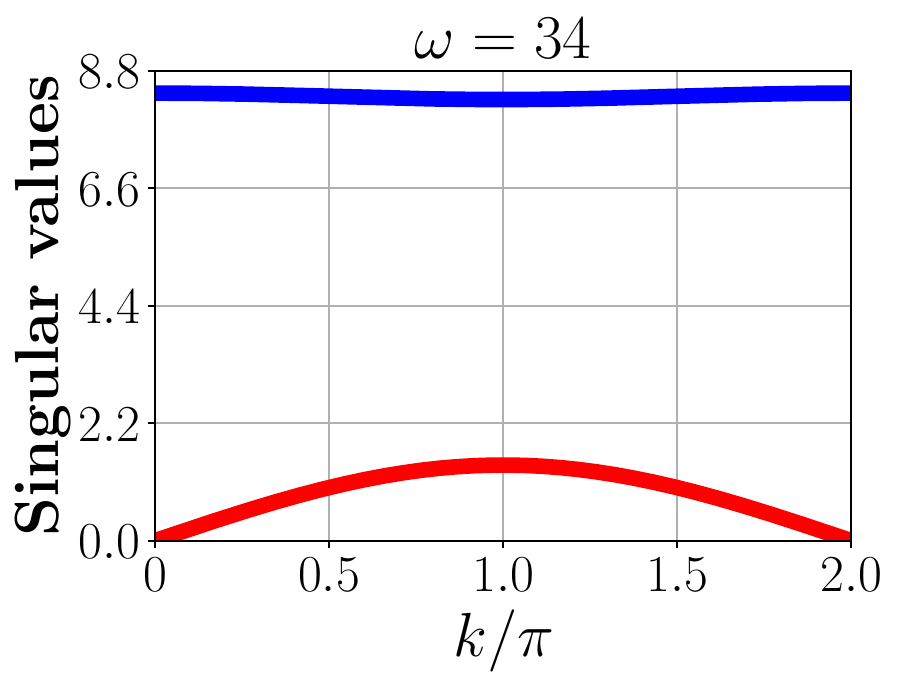} 
  \caption{}\label{HA-34} 
  \end{subfigure}
 \caption{
\emph{Singular-value spectrum of the Hermitian Hamiltonian $H^{I}(k,\omega)$.} Panels (a), (b), and (d) show the gapped spectra for $\omega = 0.5$, $1.5$, and $34$, demonstrating that $H^{I}$ remains gapped for $\omega \neq 1$. Panel (c) corresponds to $\omega = 1$, where the gap closes at $k = \pi$, marking the topological transition between the $\nu = 0$ and $\nu = 1$ phases.}\label{HA-SVD}
\end{figure}

The model I of the SSH Hamiltonian in Eq~\eqref{model} is described by the following matrix 
\begin{equation}\label{Model-1}
 H^I(\omega,k)=\begin{pmatrix}
1 + \omega & 1 + \omega{e^{-i k}}  \\
1 + e^{i k} \omega & 1 + \omega
\end{pmatrix}.  
\end{equation}
This Hermitian system exhibits a topological transition at $\omega=1$ from winding number $\nu=0$ to $\nu=1$.
To ensure the positive spectrum of $H^{I}$, we always choose $\omega>0$. 
For $\omega\neq 1$, the spectrum is gapped, and the gap closes exactly 
 at $\omega=1$ and $k=\pi$ as demonstrated in Fig.~\eqref{HA-SVD}.
 
We observe a topological transition in the NH system described by $A$ (which constructs the Hermitian system $H^I(\omega,k)$) exactly at $\omega=1$. However, in contrast to a general NH system where the topological transition is typically associated with the presence of an EP \cite{PhysRevLett.126.010401}, this transition at $\omega=1$ is characterised by a discontinuity in the eigenvalue spectrum, rather than the coalescence of eigenvalues. As a result, the topological transition in this case cannot be attributed to the presence of an EP. This interesting observation is independent of a particular choice of $A$ as demonstrated in {\bf[Appendix~\eqref{appb}]}. However, for certain choices of $V$, we do observe an exceptional point (EP), which induces a topological phase transition at the NH level, but this does not lead to any corresponding topological change at the Hermitian level, as shown in {\bf[Appendix~\eqref{appc}]}. Now we exhibit our results in detail with a particular choice of $A$. 

For $V=\sigma_x$, we obtain the NH $A$ matrix using Eq.~\eqref{SVD} as,
    \begin{equation}\label{HA-A-sigma-x}
        A=\begin{pmatrix}
    -\frac{e^{i k}a 
    \sqrt{1 + \omega - a}}
    {\sqrt{2} (e^{i k} + \omega)} &

    \frac{e^{i k} a 
    \sqrt{1 + \omega + a}}
    {\sqrt{2} (e^{i k} + \omega)} \\[10pt]

    \frac{\sqrt{1 + \omega - a}}{\sqrt{2}} & 
    \frac{\sqrt{1 + \omega + a}}{\sqrt{2}}
    \end{pmatrix},
    \end{equation} where $a=\sqrt{1 + \omega^2 + 2 \omega \cos k}$.

\begin{figure}[htbp!]
  \begin{subfigure}{0.49\columnwidth}
  \includegraphics[width=\textwidth]{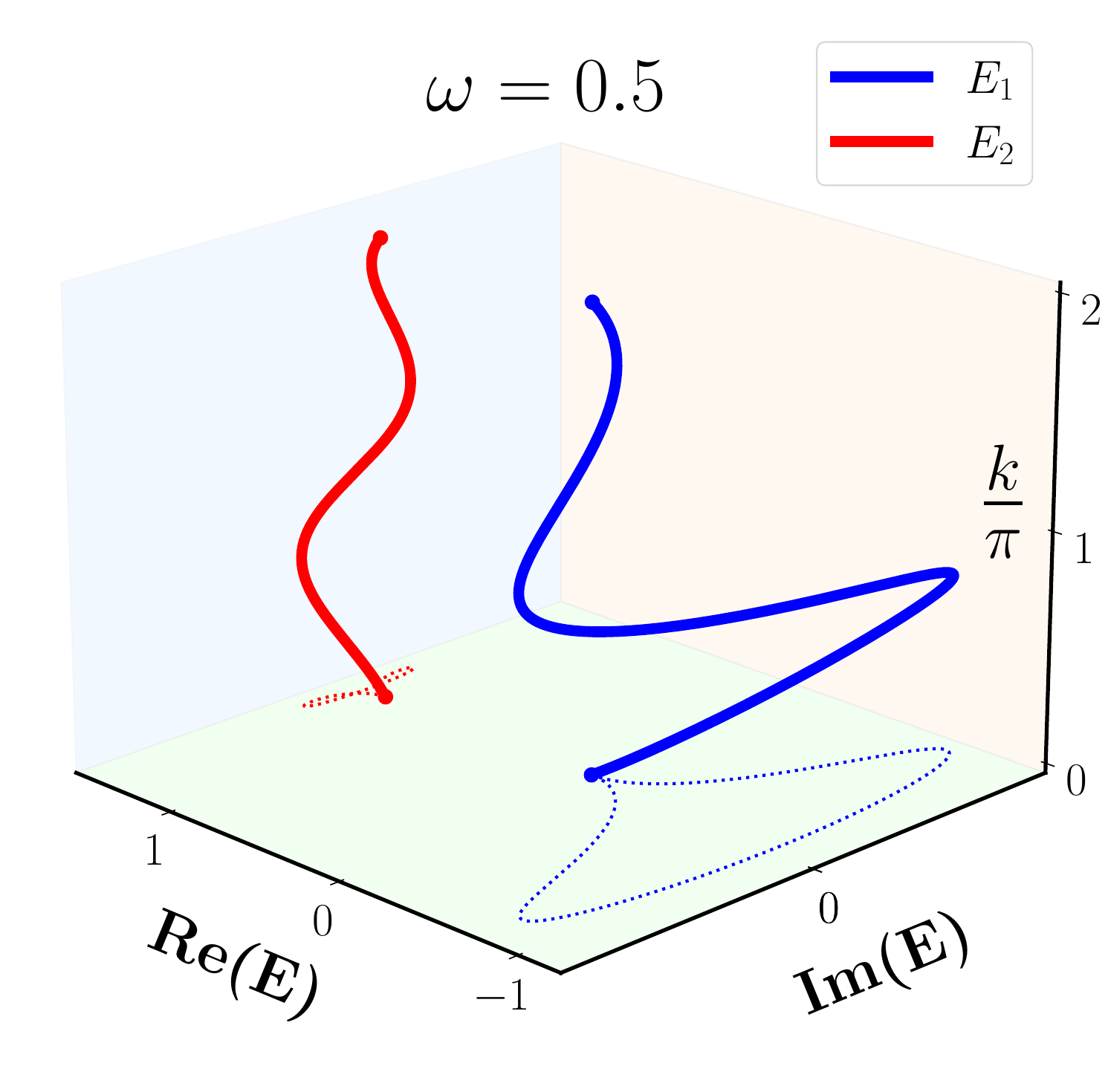}
  \caption{}\label{A11-0_5-unlink}
  \end{subfigure}
  \begin{subfigure}{0.49\columnwidth} 
  \includegraphics[width=\textwidth]{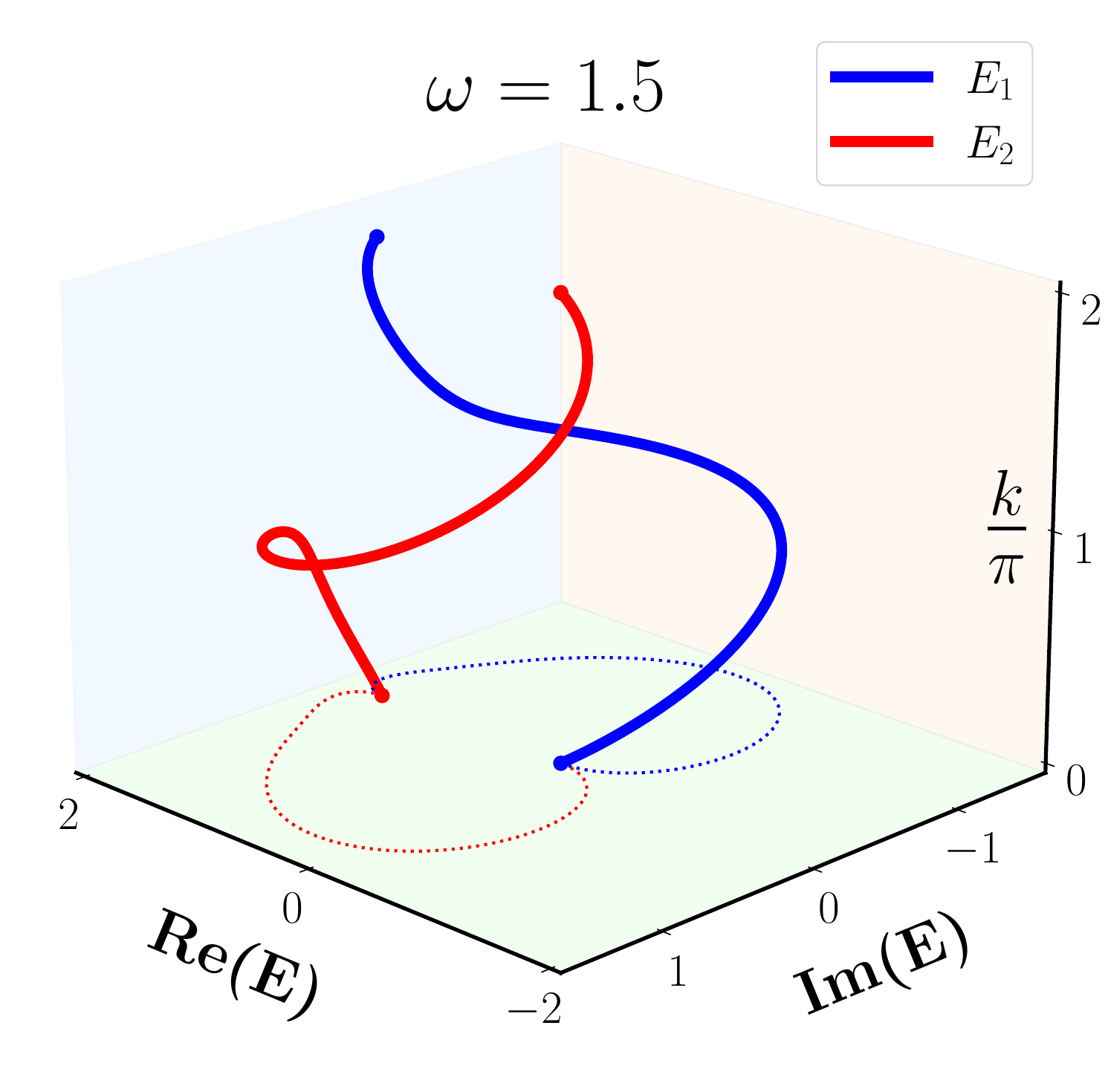} 
  \caption{}\label{A11-1_5-unknot} 
  \end{subfigure}
  \caption{
\emph{Knot structures of the NH matrix $A(\omega,k)$ for Model~I with $V=\sigma_x$}.
Panel (a) shows the unlinked knot structure for $\omega = 0.5$, while panel (b) displays the unknot for $\omega = 1.5$. These two configurations correspond to winding numbers $\nu = 0$ and $|\nu| = 1$, respectively, illustrating the change in knot topology across the transition at $\omega = 1$.}
\label{A11-HA}
  \end{figure}

The knot structures for this system are shown in the Fig.~\eqref{A11-HA} for $\omega<1$ and $\omega>1$.

We observe an unlink $(\omega<1)$ to unknot $(\omega>1)$ topological transition with winding number changing from $\nu=0$ to $|\nu|=1$ for this case. Our result is robust as long as the choices of $V$ are concerned. However, we would like to point out that for certain choices of $V$, the knot transition in $A$ happens in the reverse way, i.e., from unknot ($|\nu|=1$) to unlink ($\nu=0$). This also has been demonstrated in {\bf[Appendix~\eqref{appd}]}.
We also emphasize that although the $A$ matrix [Eq.~\eqref{HA-A-sigma-x}] appears complicated in general, in the vicinity of the knot transition point, i.e., for $k \simeq \pi$ and $\omega = 1$, the $A$ matrix can be effectively expressed in terms of a realistic NH lattice model. The details are presented in {\bf[Appendix~\eqref{appdF}]}.
\subsection{Model II}\label{MM-2}
The Hamiltonian for extended version of SSH model for this case is given by
\begin{equation}\label{Model-2}
H^{II}(\omega,k) =
\scalebox{0.8}{$
\begin{pmatrix}
3 + \omega & 1 + 2 \cos{k} + \omega e^{-2ik} \\
1 + 2 \cos{k} + \omega e^{2ik} & 3 + \omega.
\end{pmatrix}
$}.
\end{equation}

\begin{figure}[htbp!]
  \begin{subfigure}{0.49\columnwidth}
  \includegraphics[width=\textwidth]{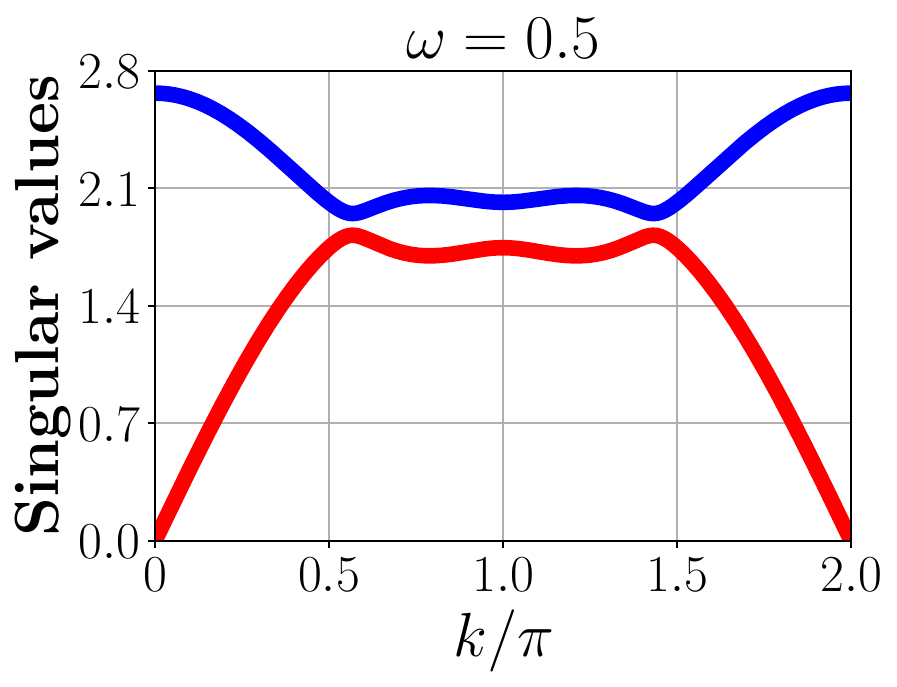}
  \caption{}\label{HB-0_5}
  \end{subfigure}
  \begin{subfigure}{0.49\columnwidth} 
  \includegraphics[width=\textwidth]{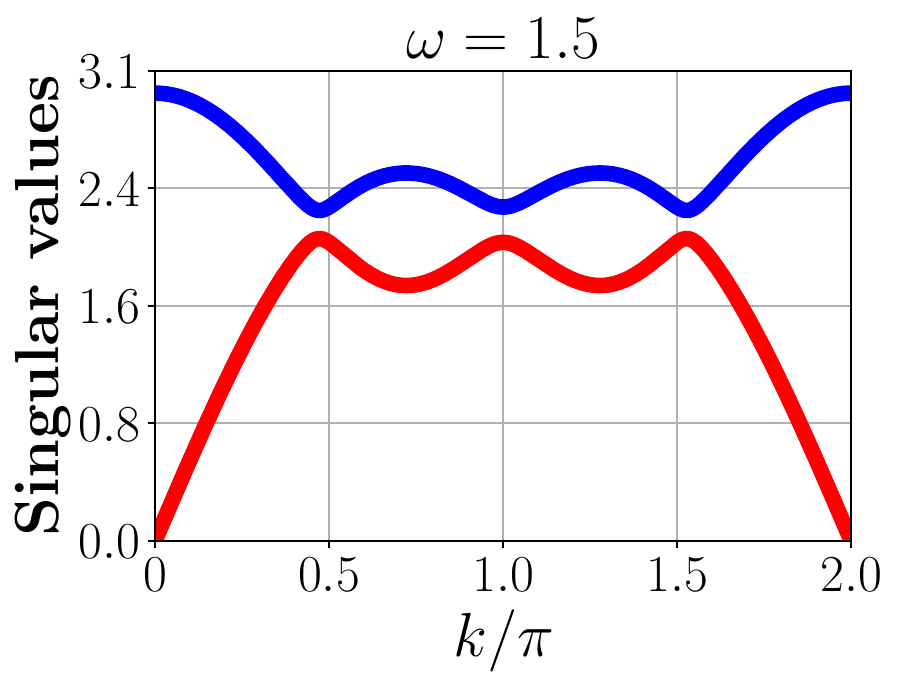} 
  \caption{}\label{HB-1_5} 
  \end{subfigure}
  \vspace{0pt}
  \begin{subfigure}{0.49\columnwidth} 
  \includegraphics[width=\textwidth]{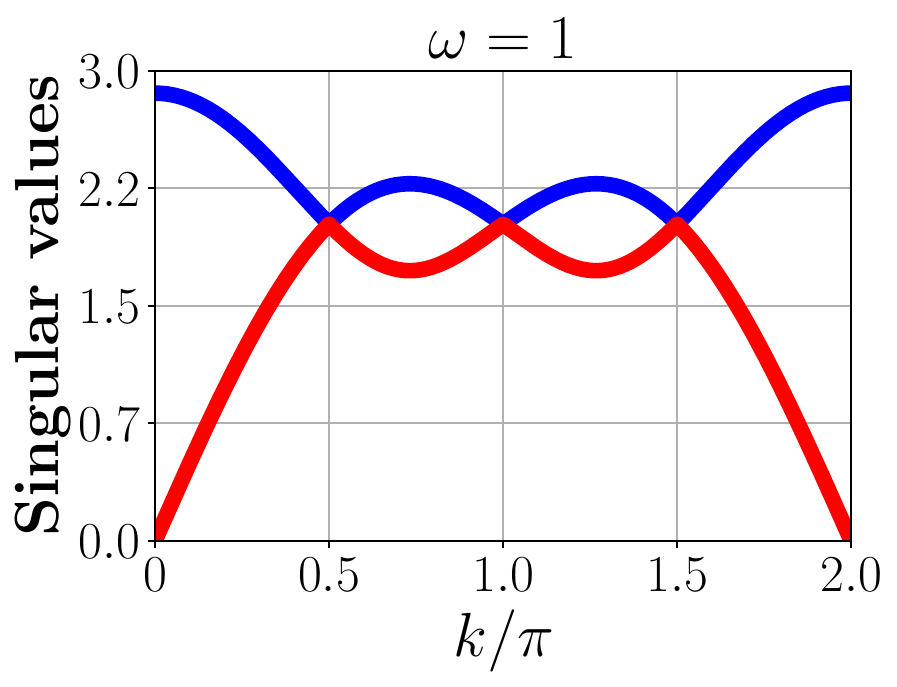} 
  \caption{}\label{HB-1} 
  \end{subfigure}
  \vspace{0pt}
  \begin{subfigure}{0.49\columnwidth} 
  \includegraphics[width=\textwidth]{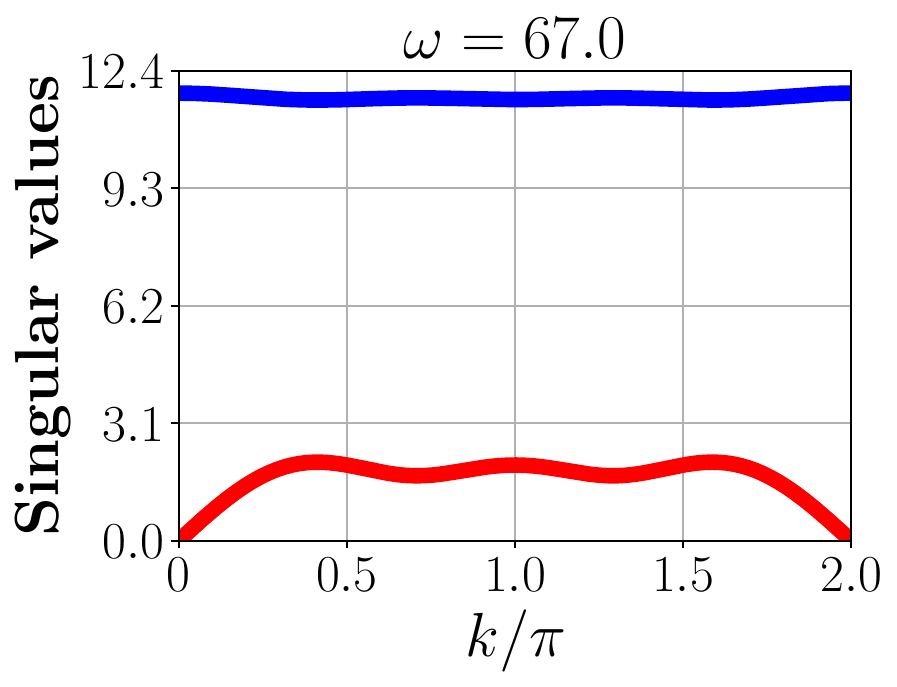} 
  \caption{}\label{HB-67} 
  \end{subfigure}
  \caption{
\emph{Singular-value spectrum of the extended SSH Hamiltonian $H^{II}(k,\omega)$}. 
Panels (a), (b), and (d) show the gapped spectra for $\omega = 0.5$, $1.5$, and $67$, 
demonstrating that $H^{II}$ remains gapped for all $\omega \neq 1$. 
Panel (c) corresponds to $\omega = 1$, where the gap closes at three points in $k$, marking the topological transition between the $\nu = 1$ and $\nu = 2$ phases. 
}
\label{HB-SVD}
  \end{figure}

This model also has a $k$ periodicity of $2\pi$ and degeneracies (gap closing) at $\omega=1$ at three different values of $k$ as shown in Fig~\eqref{HB-1}.
To ensure the positive spectrum of $H^{II}$, we once again choose $\omega>0$. 
However, in contrast to model I, where the topological transition at $\omega=1$ was from $\nu=0$ to $1$ phase, for model II, $\omega=1$ point corresponds to a topological phase transition between a winding number $\nu=1$ and $\nu=2$. Fig.~\eqref{HB-SVD} shows the plots indicating the gap closing and no-gap spectra for different values of $\omega$.

\begin{figure}[htbp!]
  \begin{subfigure}{0.49\columnwidth}
  \includegraphics[width=\textwidth]{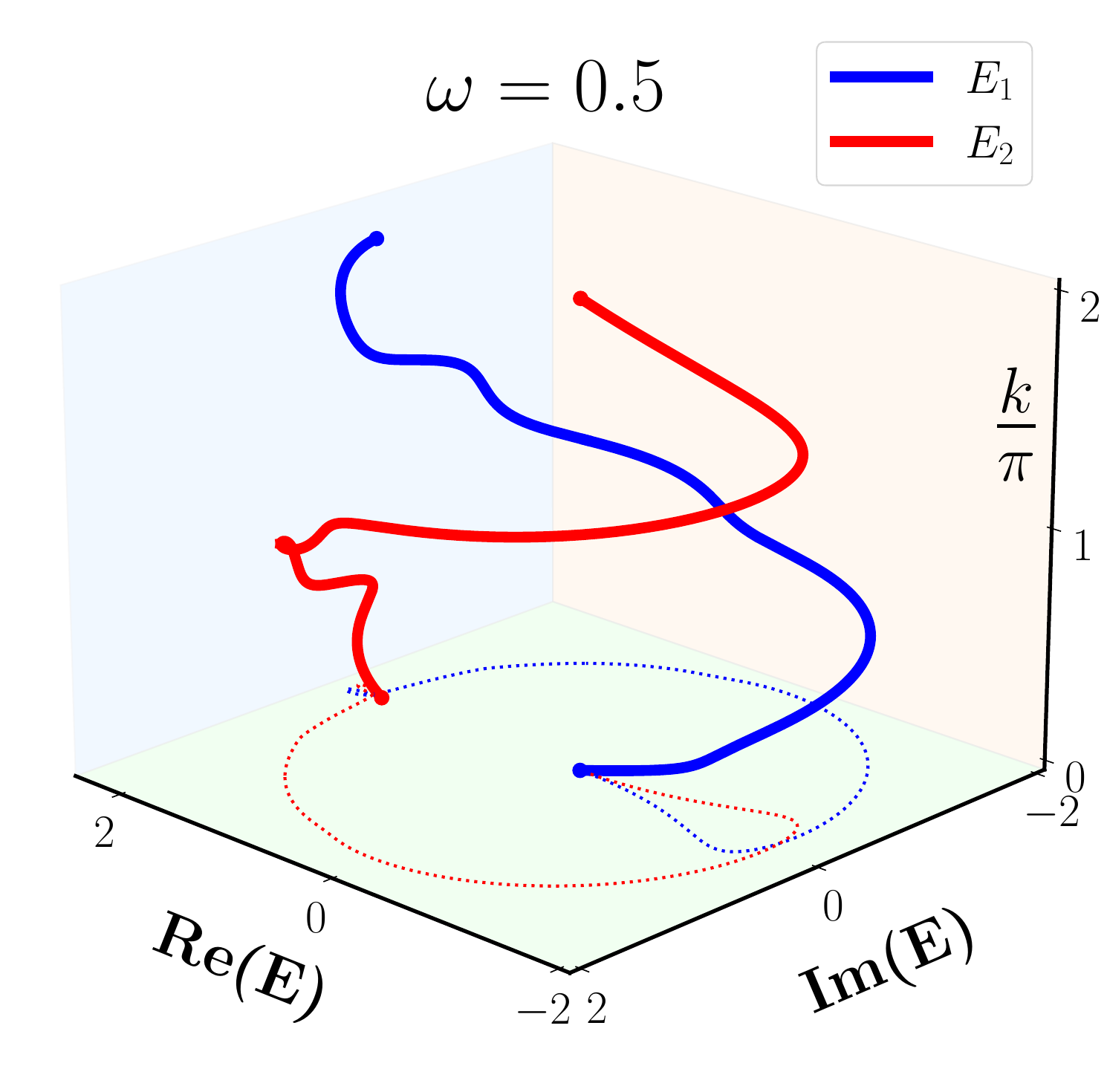}
  \caption{}\label{A11-0_5-unknot}
  \end{subfigure}
  \begin{subfigure}{0.49\columnwidth} 
  \includegraphics[width=\textwidth]{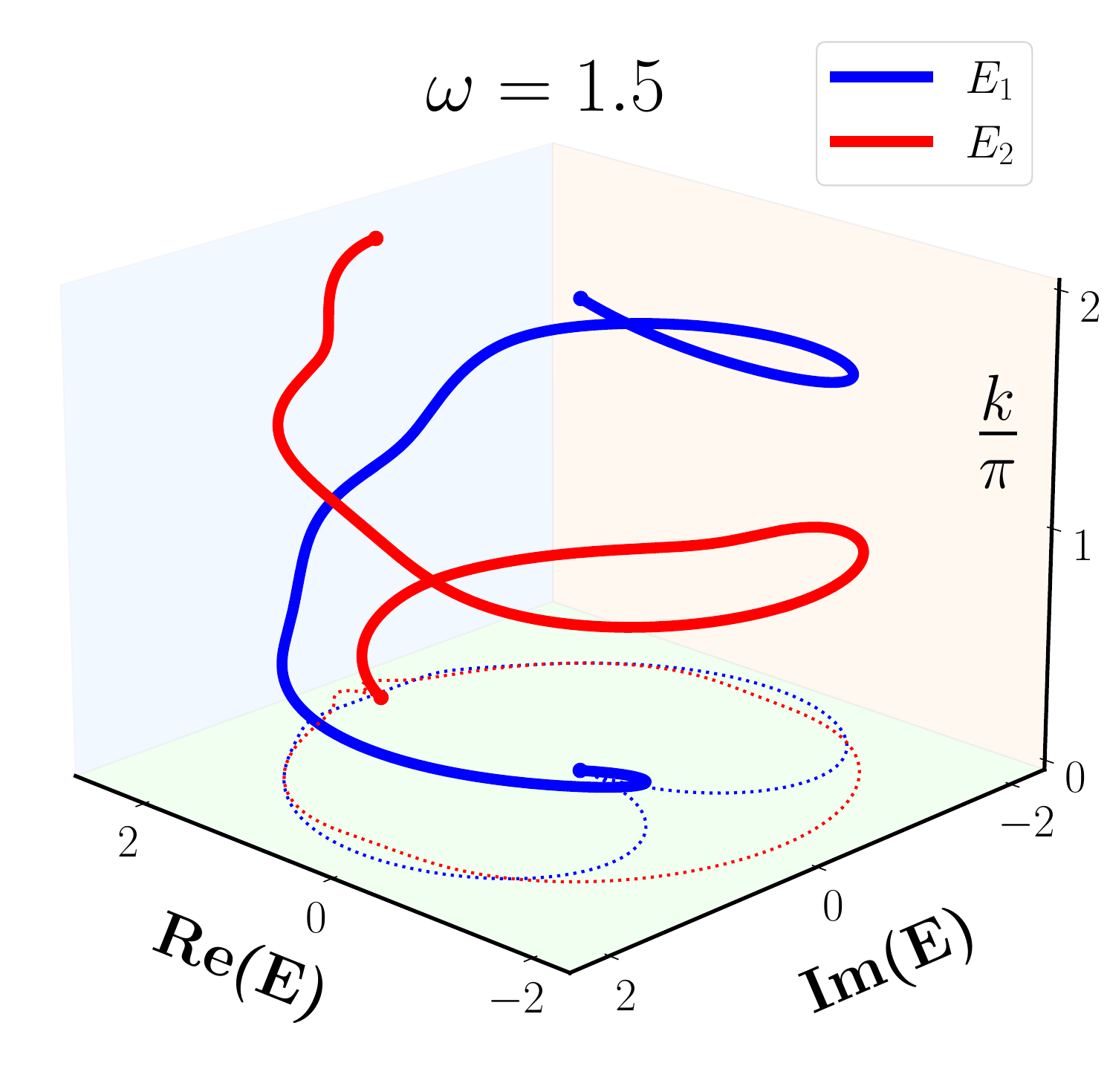} 
  \caption{}\label{A11-1_5-hopflink} 
  \end{subfigure}
  \caption{
\emph{Knot structures of the NH matrix $A(\omega,k)$ for Model~II with $V=\sigma_x$}. 
Panel (a) shows the unknot configuration for $\omega = 0.5$, while panel (b) displays the 
Hopf-link structure for $\omega = 1.5$. These correspond to winding numbers $|\nu| = 1$ 
and $|\nu| = 2$, respectively, illustrating the knot-topology change across the transition at $\omega = 1$.
}
\label{A1-HB}
  \end{figure}

We consider the same choice of $V$ for NH system in this model too. Spectrum discontinuity at $\omega=1$ in this NH system too indicates a topological phase transition {\bf[Appendix~\eqref{appb}]}.
In this case we observe unknot $(|\nu|=1)$ to hopflink $(|\nu|=2)$ knot transition as graphically shown in Fig.~\eqref{A1-HB}.

The NH system $A$, with $V=\sigma_x$ for this model is represented by the matrix
    \begin{equation}\label{HB-A-sigma-x}
        A=\begin{pmatrix}
-\frac{a'\sqrt{
\left(3 + \omega - a'\right)}}{\sqrt{2} (1 + e^{2i k} \omega + 2 \cos k)}
& 
\frac{a'\sqrt{ 
\left(3 + \omega + a'\right)}}{\sqrt{2} (1 + e^{2i k} \omega + 2 \cos k)} \\[10pt]
\frac{\sqrt{3 + \omega - a'}}{\sqrt{2}} 
& 
\frac{\sqrt{3 + \omega + a'}}{\sqrt{2}}
\end{pmatrix},
\end{equation}
where 
\begin{equation}
a'=\sqrt{3 + \omega^2 + 2\left(\left(2+\omega\right)\cos k + \left(1 + \omega\right) \cos 2k + \omega \cos 3k\right)}\nonumber.
\end{equation}
The knot structures for $\omega<1$ amd $\omega>1$ for this system are shown in Fig.~\eqref{A1-HB}

A similar unknot-hopflink knot transition is observed for a different choice of $V$, which has been demonstrated in {\bf[Appendix~\eqref{appd}]}. 
\begin{figure}[htbp!]
    \centering \includegraphics[width=0.5\textwidth]{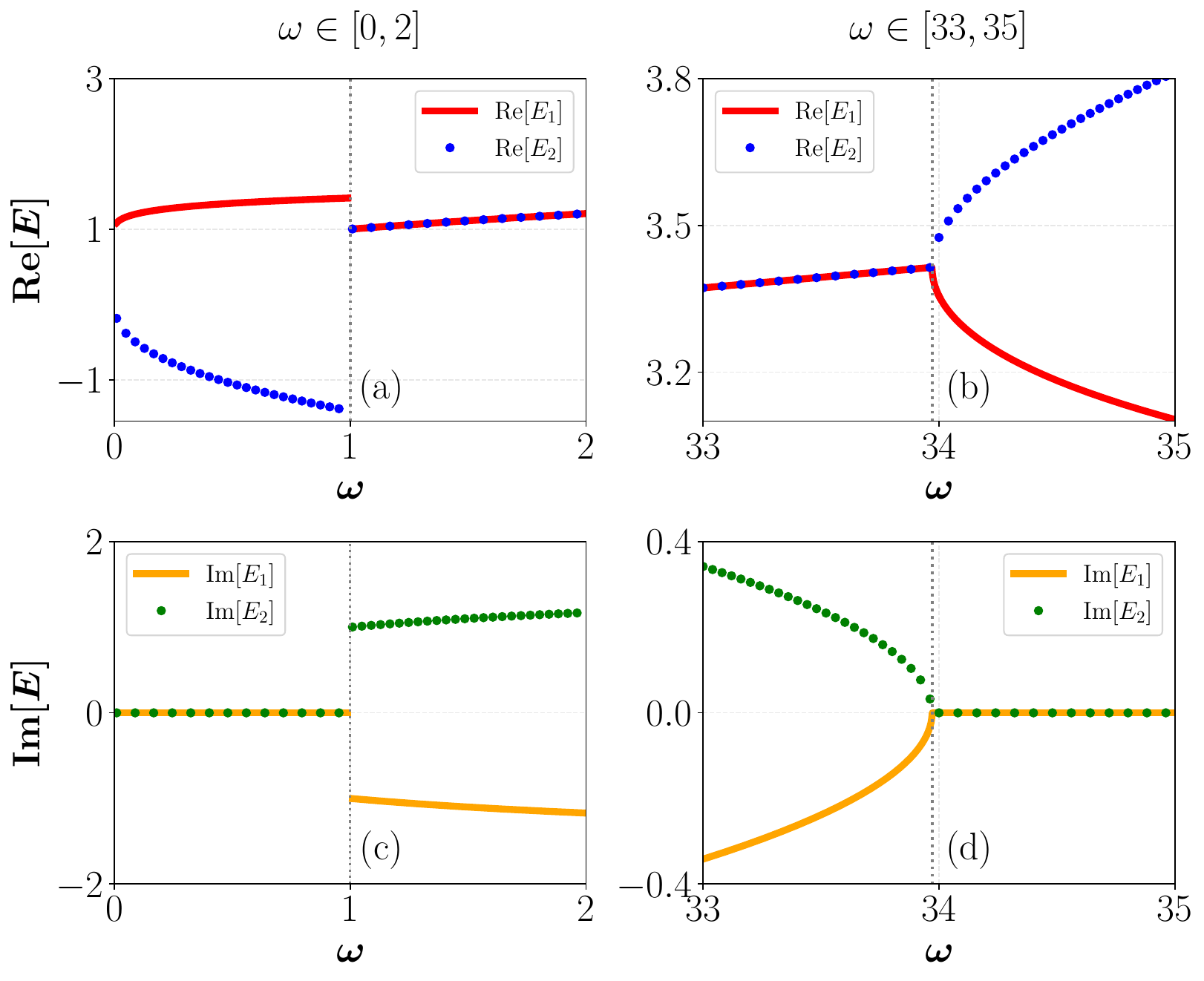}
    \caption{
\emph{Real and imaginary parts of the eigenvalues of $A(\omega,k)$ at $k=\pi$ for Model~I}. 
Panels (a) and (c) show the discontinuous jump at $\omega = 1$, indicating the first-order 
knot transition. Panels (b) and (d) show the behavior near the exceptional point at 
$\omega \simeq 34$, where the transition occurs without any discontinuity.
}\label{c-dc}
\end{figure}

\section{First order knot transition:} We now have enough evidence to support that the topological transition in $H$ will have an imprint in $A$ in the form of a knot transition. Certainly, this knot transition can not be accompanied by an EP (as discussed in the formalism section). We investigate it further and find that the real and imaginary part of the eigenvalues show discontinuity at the transition point $\omega=1$; in contrast to another knot transition point at $\omega\simeq34$ for model I and at $\omega\simeq 67$ for model II (see {\bf[Appendix~\eqref{appc}]}), which are accompanied by EP, and the real and imaginary part of the eigenvalues do not show any discontinuity (see Fig.\eqref{c-dc} and Fig.\eqref{c-ec}). We refer to this new type of knot transition at $\omega=1$, i.e., not accompanied by an EP, as {\it{first order knot transition}}.

 \section{Discussions}\label{discussions} 
Our main goal in this manuscript was to establish a connection between the knot topology of complex eigenvalues of an NH matrix and its real singular values. We framed the problem a bit differently, i.e., is there a correlation between the topological transition in the Hermitian model $H$ and the knot topology of complex eigenvalues of the underlying NH Hamiltonian $A$, such that $H=AA^{\dag}$.  We find that indeed there is a correlation whenever the Hermitian $H$ goes through a topological transition in terms of the winding number, the eigenvalues of NH $A$ (provided the $k$ periodicity is the same as $H$) also go through a transition between two distinct knot topologies. We validate these claims by choosing different NH matrices. This knot transition is not accompanied by an EP; instead, the eigenvalues of the NH matrices show a discontinuity at the transition point, which we refer to as {\it{first order knot transition}}.
Interestingly, we also find an example in the same model that shows a knot transition between two different knot topologies, accompanied by an EP, which does not show a topological transition in Hermitian $H$. To the best of our knowledge, this particular approach provides a complementary perspective on connecting the physics of topology between Hermitian and NH systems, differing from earlier studies. Our findings suggest that signatures of Hermitian topology can persist in the corresponding NH construction, whereas the converse correspondence does not generally occur. This should motivate us to construct NH models out of many exotic Hermitian systems in higher dimensions, which show rich phases such as quantum-hall systems~\cite{von202040}, higher-order topological insulators~\cite{schindler2018higher,zhang2009topological}, or semi-metals~\cite{lv2015experimental}. The first step will be to theoretically construct NH systems whose Hermitian counterpart shows interesting exotic phases, then the theoretical question one can ask is whether such Hermitian exotic phases keep some interesting imprint in the underlying NH system or not. The next step will be to probe such NH phases using a photonics experimental platform, where the NH system can be visualized efficiently~\cite{wang2021topological}. It is also worth mentioning that a recent work has established a connection between exceptional points (EPs) of the non-Hermitian transfer matrix of quadratic finite-range Hermitian Hamiltonians and van Hove singularities (VHSs), where the density of states diverges~\cite{PhysRevB.111.L041405}. In future studies, it would be interesting to explore whether an analogous relationship exists for our knot transition points, despite the absence of EPs.

 \section{Acknowledgments}
 G.H. acknowledges  the fruitful discussion with  Aditya Dwivedi on Knot Theory and  the UGC-SRF. R.M. acknowledges the DST-Inspire fellowship by the Department of Science and Technology, Government of India, SERB start-up grant (SRG/2021/002152). B.P.M. acknowledges the PDF grant of IOE, BHU for the year 2025-26. 
 
 {\bf DATA AVAILABILITY}
 
The data that support the findings of this article are not
publicly available. The data are available from the authors
upon reasonable request. 
\newpage

\bibliography{ref_3}
\onecolumngrid
\newpage

\appendix
\section*{Appendix}
In the main text, we consider only one choice of $V$ for each of models I and II to demonstrate the results. In this Appendix, we show that our conclusions are robust by analyzing a $k$-dependent $V$ matrix, the discontinuity at the transition point, the presence of an additional transition point, and additional examples of $V$ for both models I and II. We further demonstrate that the presence of a topological transition point in the NH system does not necessarily imply a corresponding transition in its Hermitian counterpart, i.e., the converse of our result is not true. Finally, we have shown an effective real space NH description of $A$.

\section{$k$-dependent $V$ matrix}\label{appa}
In the main part of the paper, we have shown that a \( k \)-independent traceless matrix \( V \) is responsible for preserving a specific pair of topological transitions. Specifically, if a Hamiltonian at the Hermitian level is associated with a topological transition from winding number 0 to 1, then a \( k \)-independent traceless \( V \) ensures that transition occurs from  $\nu=0$ phase to $|\nu|=1$ phase; which in turn implies transition between unlink and unknot. Alternatively, if the Hermitian Hamiltonian is associated with a transition from $\nu=1$ to $\nu=2$, then again, a traceless, \( k \)-independent \( V \) ensures transition from unknot to hopflink, as these correspond to the winding number values of 1 and 2 at the NH level. Differences arise when we consider a \( k \)-dependent matrix \( V \), for which the associated knot structures can become random at the NH level, as illustrated in the example below.

\begin{figure}[htbp!]
  \begin{subfigure}{0.3\columnwidth}
  \includegraphics[width=\textwidth]{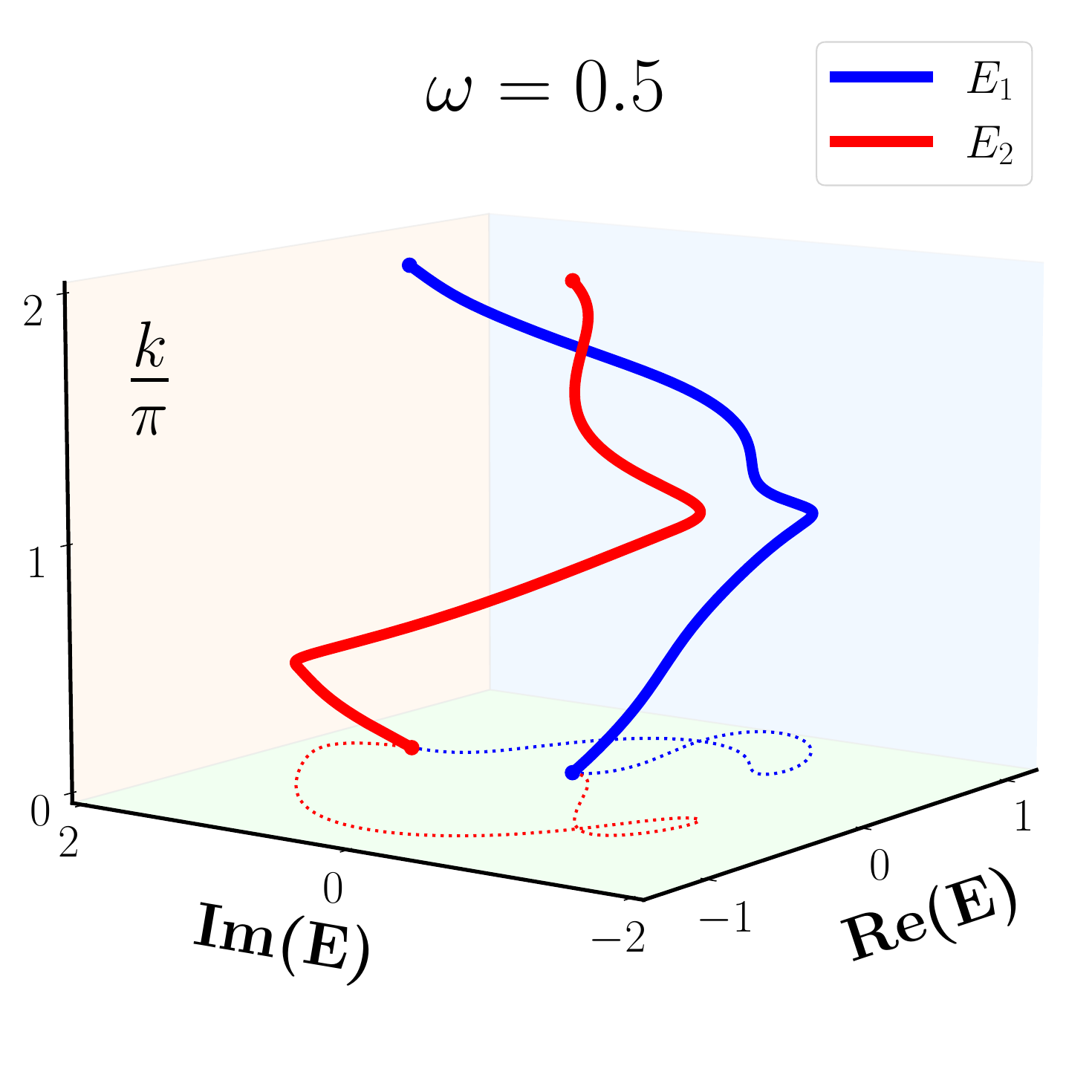}
  \caption{}\label{A33-0_5}
  \end{subfigure}
  \begin{subfigure}{0.3\columnwidth} 
  \includegraphics[width=\textwidth]{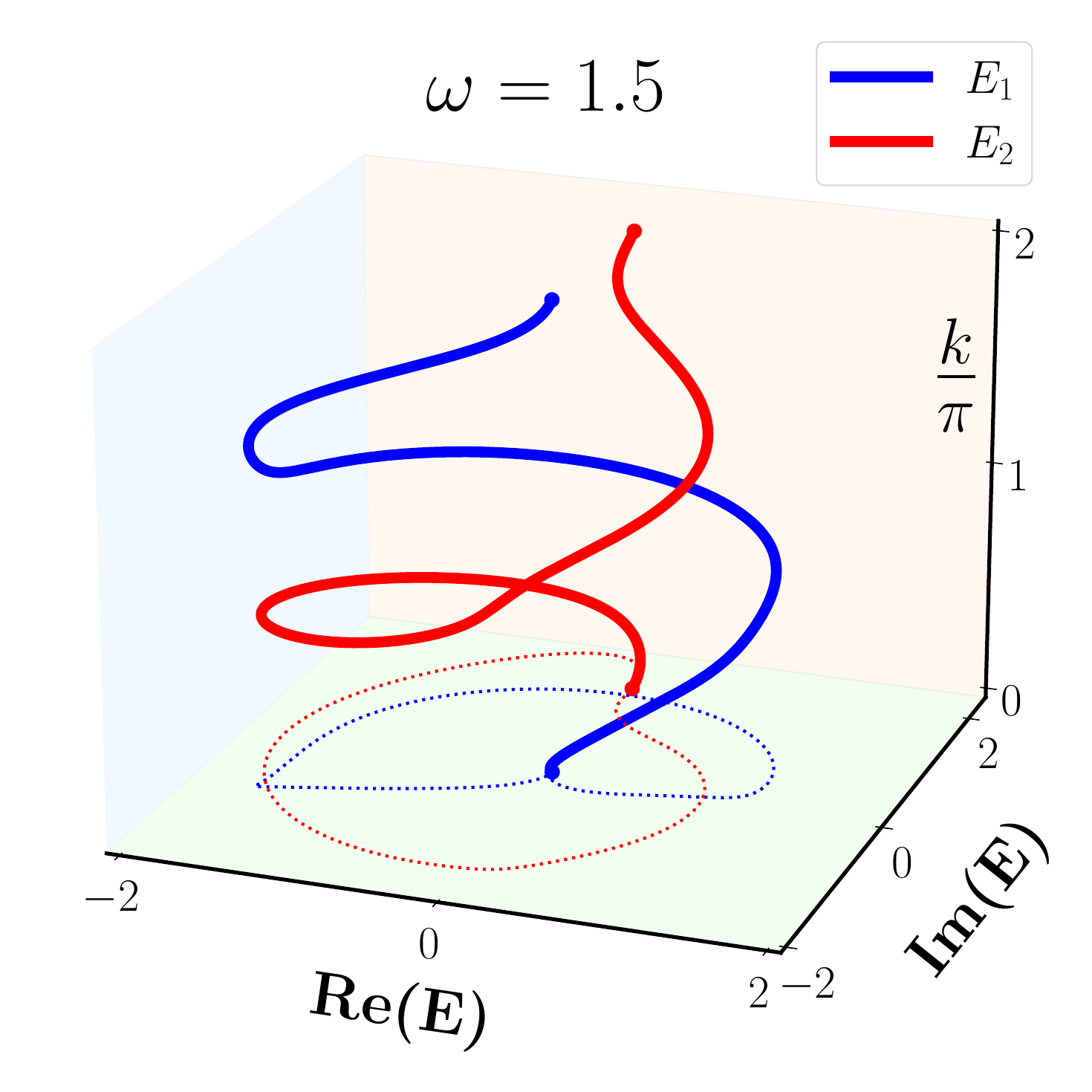} 
  \caption{}\label{A33-1_5} 
  \end{subfigure}
  \caption{
\emph{Knot structures of $A(\omega,k)$ for Model~I with the $k$-dependent matrix $V(k)$}. 
Panel (a) shows the unknot for $\omega = 0.5$, while panel (b) displays the Hopf-link for 
$\omega = 1.5$. The use of a $k$-dependent $V$ changes the resulting knot types compared 
to the traceless, $k$-independent choices.
}\label{A33-HA}
\end{figure}

We consider the following $k$ dependent $V$ and use it to construct the $A$ matrix for both the models and analyse their knot structures:
\begin{equation}\label{k-dependent}
    V=\begin{pmatrix}
- i \sin(k) & - i e^{-i k} \cos(k) \\
- i \cos(k) & i\sin(k) e^{-ik}
\end{pmatrix}.
\end{equation}
For the first model, the corresponding NH $A$ is obtained as,
\begin{equation}
    A=\begin{pmatrix}
\displaystyle
- \frac{i e^{i k} a}{\sqrt{2}(e^{i k} + \omega)} 
\left( e^{i k} \cos(k) N - M \sin(k) \right)
&
\displaystyle
\frac{i e^{i k} a}{\sqrt{2}(e^{i k} + \omega)} 
\left( \cos(k) M + e^{i k} N \sin(k) \right)
\\[1.2em]
\displaystyle
\frac{i}{\sqrt{2}} \left( e^{i k} \cos(k) N + M \sin(k) \right)
&
\displaystyle
\frac{i}{\sqrt{2}} \left( \cos(k) M - e^{i k} N \sin(k) \right)
\end{pmatrix},
\end{equation}
where $a=\sqrt{1 + \omega^2 + 2 \omega \cos k}$, $M=\sqrt{1 + \omega + a}$ and $N=\sqrt{1 + \omega - a}$. For this form of $A$, the knot structures obtained are shown in the Fig.~\eqref{A33-HA}.

\begin{figure}[htbp!]
  \begin{subfigure}{0.3\columnwidth}
  \includegraphics[width=\textwidth]{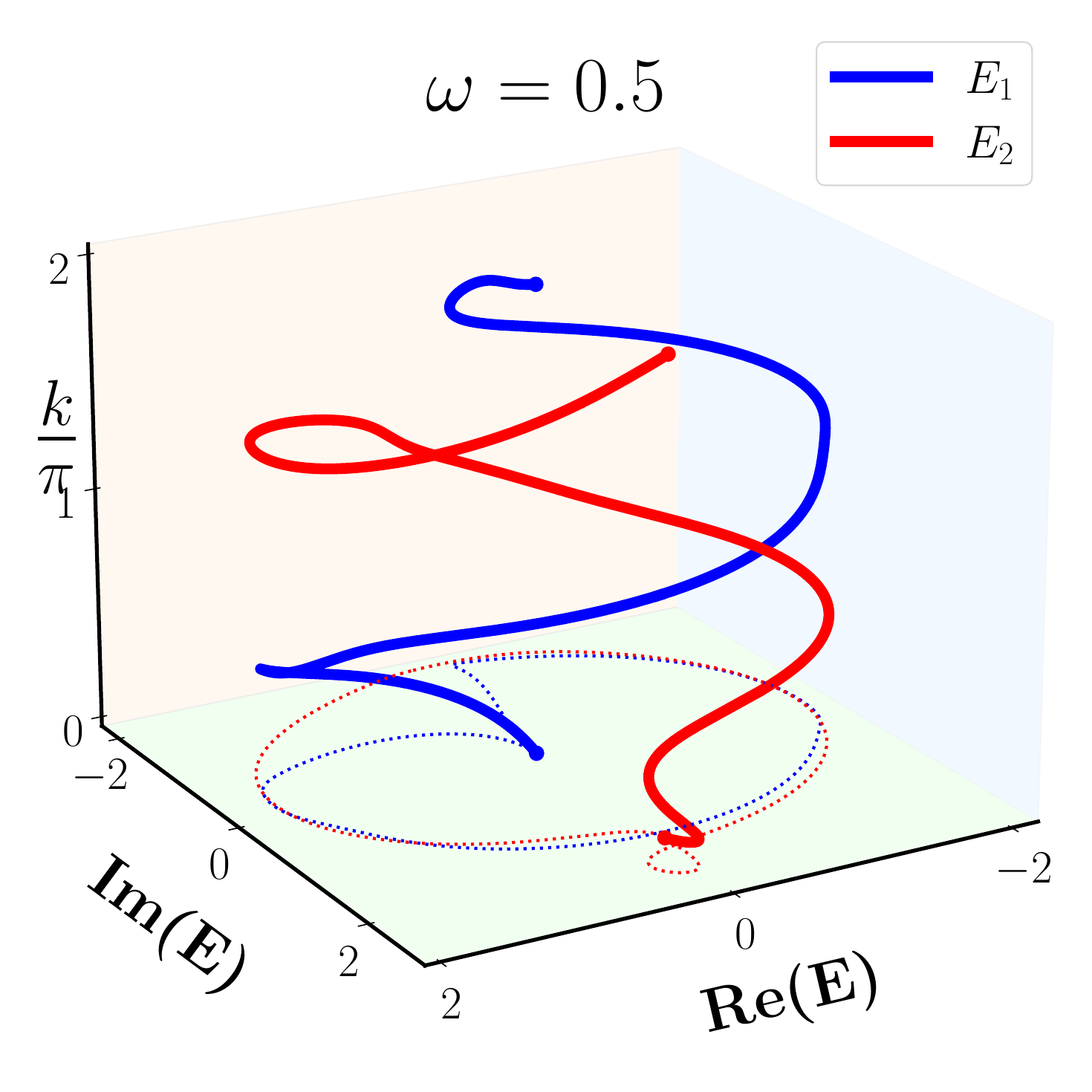}
  \caption{}\label{M2-0_5}
  \end{subfigure}
  \begin{subfigure}{0.3\columnwidth} 
  \includegraphics[width=\textwidth]{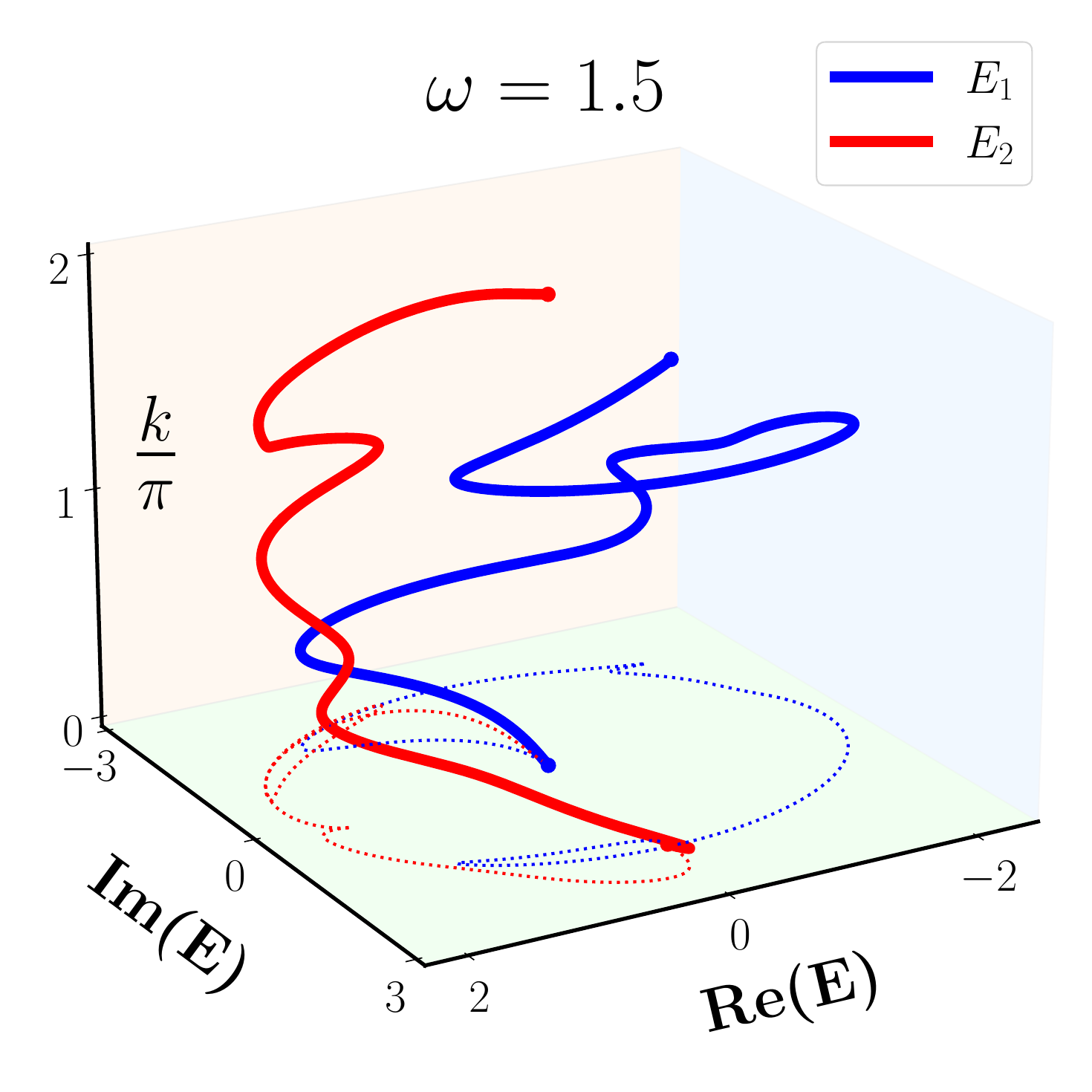} 
  \caption{}\label{M2-1_5} 
  \end{subfigure}
  \caption{
\emph{Knot structures of $A(\omega,k)$ for Model~II with the $k$-dependent matrix $V(k)$}. 
Panel (a) shows the hopflink for $\omega = 0.5$, while panel (b) shows the unknot for 
$\omega = 1.5$. As in Model~I, the $k$-dependence of $V$ leads to knot configurations 
distinct from those obtained with traceless, $k$-independent choices.
}\label{M2-HB}
\end{figure}

Similarly for model II, the NH $A$ is obtained as:
\begin{equation}
A=\begin{pmatrix}
\dfrac{ ia'\left(M'\sin k
-e^{ik}  N' \cos k
\right)}{
\sqrt{2} (1 + e^{2ik} \omega + 2 \cos k)
}
&

\dfrac{ia'\left(
M'\cos k
+ N'e^{ik}\sin k
\right)}{
\sqrt{2} (1 + e^{2ik} \omega + 2 \cos k)
}
\\ \\

\dfrac{
i}{
\sqrt{2}
} \left(M' \sin k + e^{ik}N' \cos k
\right)
&

\dfrac{
i}{
\sqrt{2}
}\left( M'\cos k
- N'e^{ik}\sin k
\right)
\end{pmatrix}
\end{equation}
where $a' = \sqrt{
3 + \omega^2 + 2 (2 + \omega) \cos k + 2 (1 + \omega) \cos 2k + 2 \omega \cos 3k
}$, $M'=\sqrt{3 + \omega + a'}$ and $N'=\sqrt{3 + \omega - a'}$. The knot structures for this system are shown in Fig.\eqref{M2-HB}.

\begin{figure}[tb]
    \centering
    \includegraphics[width=0.3\linewidth]{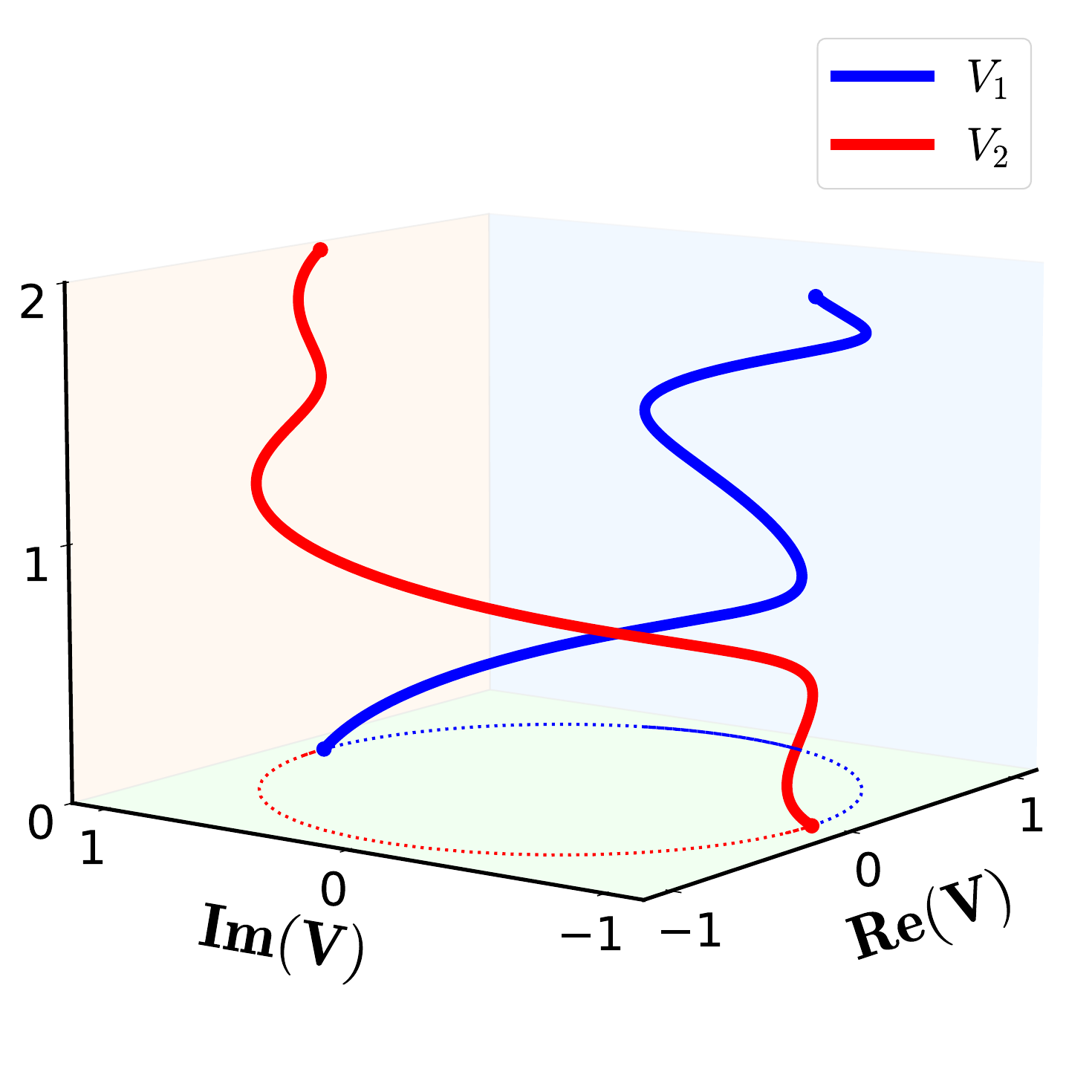}
    \caption{\emph{Eigenvalue spectrum of the unitary matrix $V(k)$}, demonstrating that $V(k)$ itself exhibits a knot structure in the complex plane as $k$ is varied.}
    \label{FVp}
\end{figure}

We observe that the set of knots in Fig.~\eqref{A33-HA} represents the unknot and the Hopf link with \( |\nu| = 1 \) and \( |\nu| = 2 \), respectively, which contrasts with the configurations shown in Figs.~\eqref{A11-HA} and \eqref{A55-HA}. For different forms of \( k \)-dependent \( V \), the set of knots will vary. Similar deviation can be seen in the case of the second model as well, which is evident from the fact that Fig.\eqref{M2-HB} shows different nature of knot transition from that of Fg.\eqref{A1-HB}. The reason behind this deviation of knot transition from the earlier cases is due to the fact that $V(k)$ itself possesses a knot which is shown in Fig.~\eqref{FVp}. As a result, the knot transition observed in $A$ will now differ from the scenario in which $V$ is momentum independent and does not possess any knot of its own. The key point we wish to emphasize is that, regardless of the specific form of \( V \), a topological transition occurs at \( \omega = 1 \) in the NH spectrum. The distinction lies in the fact that a traceless, \( k \)-independent \( V \) preserves the specific form of the knots, whereas \( k \)-dependent \( V \) leads to transition with different knots.

\begin{figure}[htbp!]
  \centering
  \begin{subfigure}{0.3\columnwidth}
    \centering
    \includegraphics[width=\textwidth]{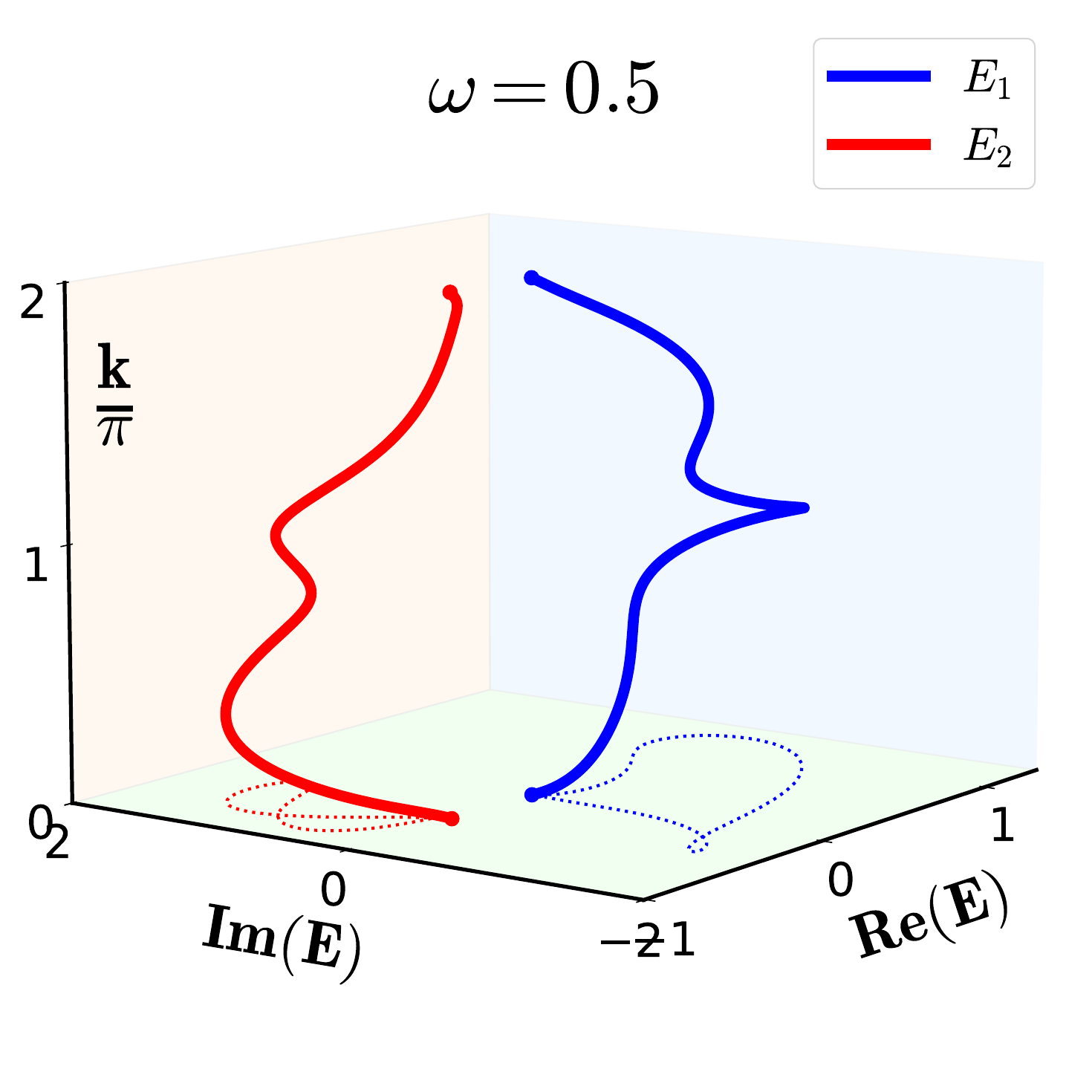}
    \caption{}
    \label{G1a}
  \end{subfigure}
  \begin{subfigure}{0.3\columnwidth}
    \centering
    \includegraphics[width=\textwidth]{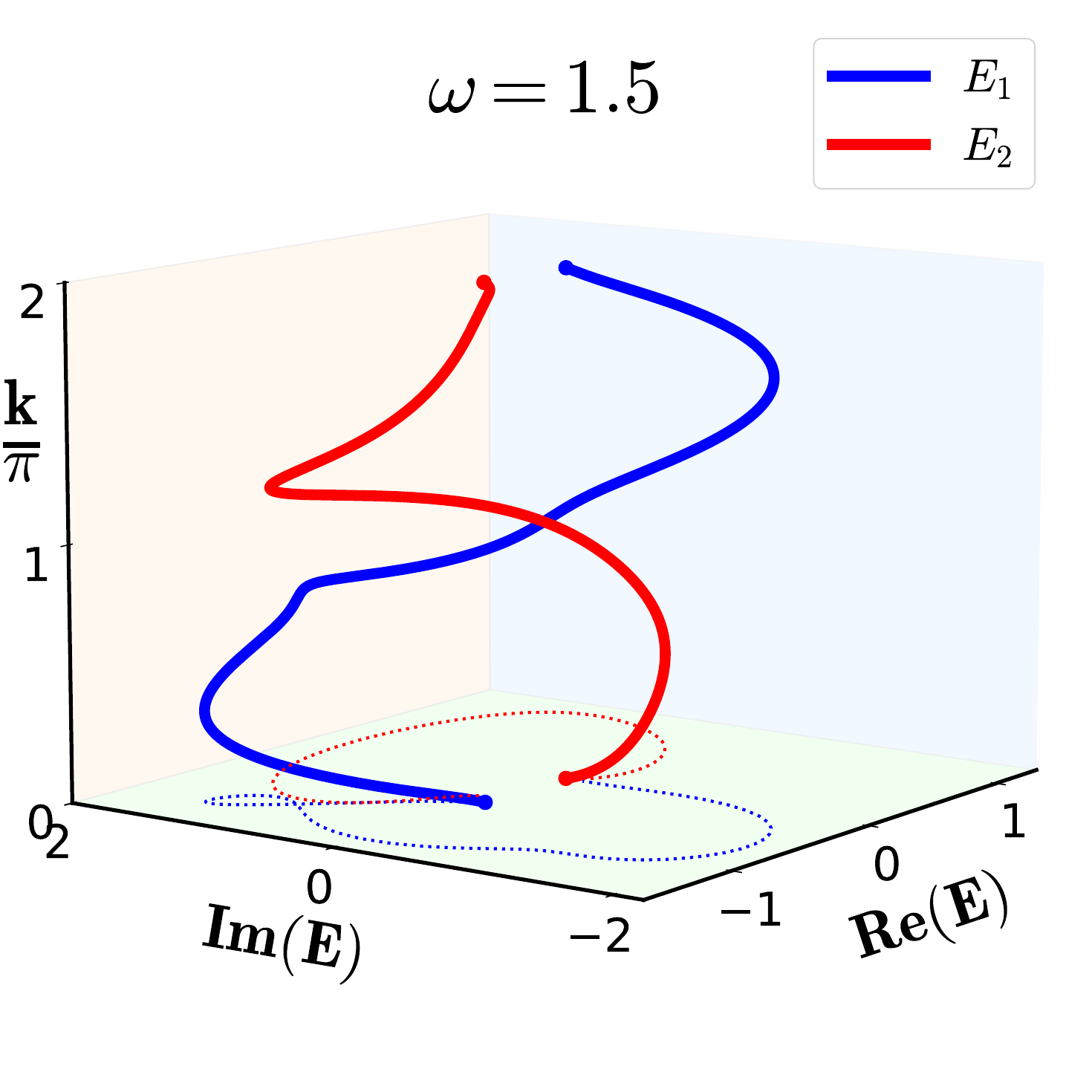}
    \caption{}
    \label{G1b}
  \end{subfigure}

  \caption{\emph{Knot structures of $A'(\omega,k)$} for Model~I, obtained using $\mathcal{U}$ with parameter values
$\varphi = \frac{\pi}{2}$, $\alpha = \frac{\pi}{3}$, $\beta = \frac{\pi}{6}$, and $\theta = k + \frac{\pi}{4}$.
Panel~(a) shows an unlink at $\omega = 0.5$, whereas panel~(b) displays an unknot at $\omega = 1.5$.}
\label{G1}

\end{figure}

We further illustrate the robustness of the knot transition under different unitary gauge choices within the SVD construction. In particular, we consider a general unitary matrix of the form
\[
\mathcal{U} = e^{i\varphi/2}
\begin{pmatrix}
e^{i\alpha}\cos\theta & e^{i\beta}\sin\theta \\
-\,e^{-i\beta}\sin\theta & e^{-i\alpha}\cos\theta
\end{pmatrix},
\]
where, $\varphi$, $\alpha$, $\beta$, $\theta$ are arbitrary parameters. With this $\mathcal{U}$ we then construct the NH matrix \(A' = A \mathcal{U}\) which continues to satisfy \(H = A' A'^\dagger\). Now making sure that the chosen parameters of $\mathcal{U}$ preserve the periodicity of the parent Hermitian Hamiltonian, we observe that the spectrum of \(A'\) exhibits a discontinuity at the same values of the parameters $(\omega,k)$ as the original matrix \(A\), and the corresponding knot structure undergoes a transition across this point Fig.~\eqref{G1}. This demonstrates that the appearance of the discontinuity and the associated knot transition persist across different unitary representatives within the gauge-related family of NH matrices generated by the SVD construction.


\section{Discontinuity at the transition point}\label{appb}

\begin{figure}[htbp!]
  \begin{subfigure}{0.3\columnwidth}
  \includegraphics[width=\textwidth]{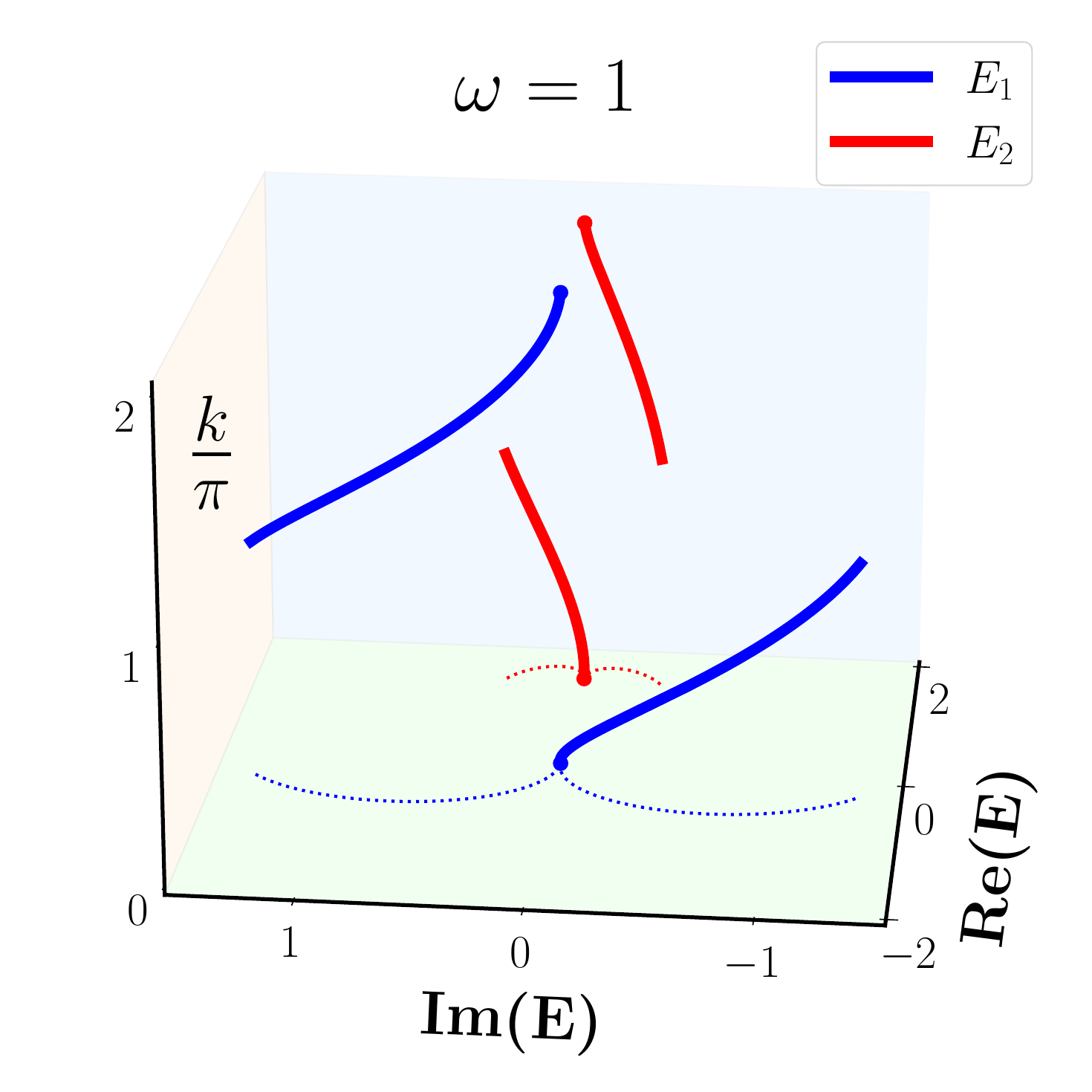}
  \caption{}\label{A11-HA-1}
  \end{subfigure}
  \begin{subfigure}{0.3\columnwidth} 
  \includegraphics[width=\textwidth]{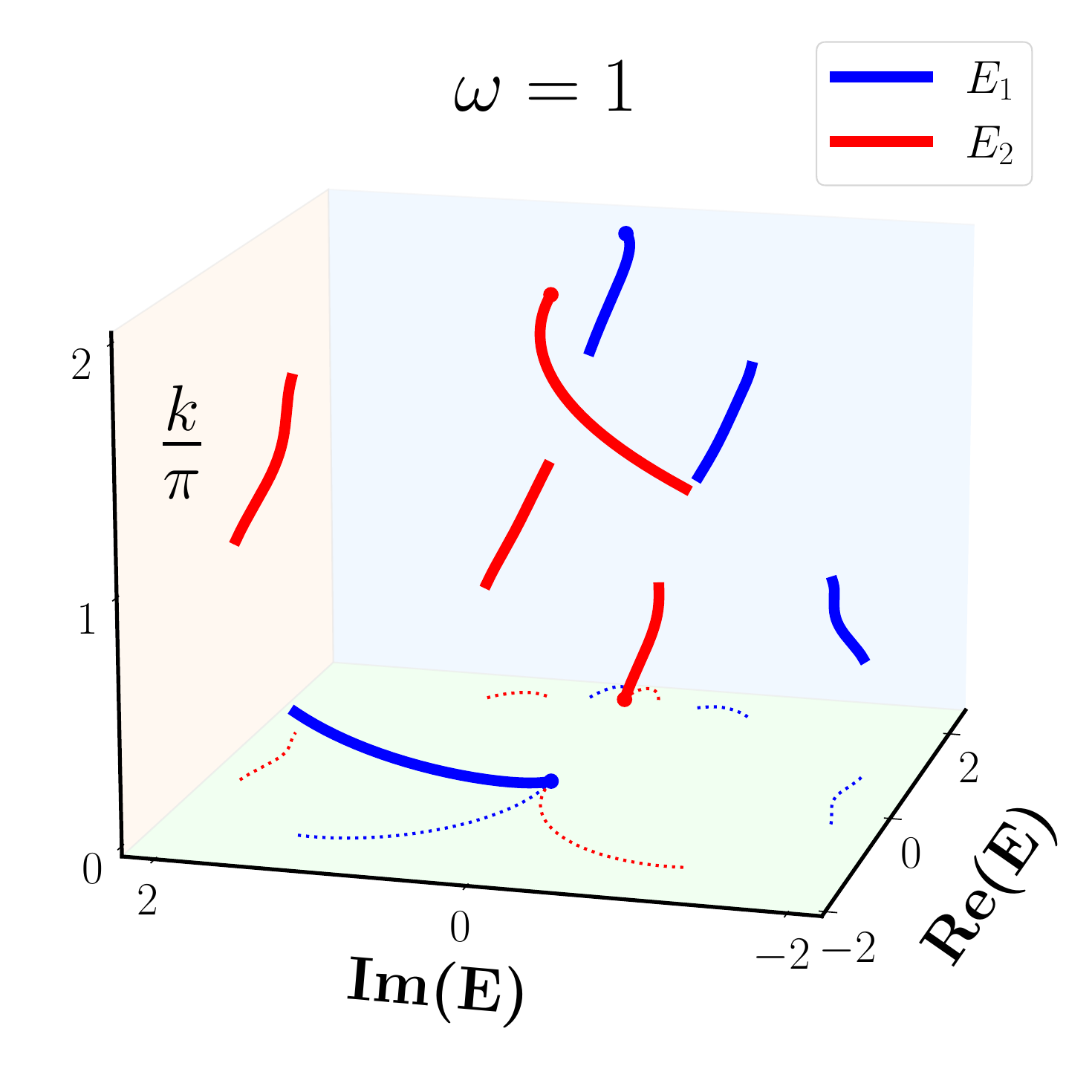} 
  \caption{}\label{A11-HB-1} 
  \end{subfigure}
  \caption{
\emph{Spectral discontinuity of the eigenvalues of $A(\omega,k)$ at the transition point $\omega = 1$}. 
Panels (a) and (b) show the discontinuous jump in the eigenvalues at $k = \pi$ for 
Model~I and Model~II, respectively, characterizing the first-order knot transition 
that occurs without an exceptional point.
}
\label{A11-HAHB-1}
\end{figure}
We have demonstrated that the NH choices of \( A \) associated with our Hermitian SSH models exhibit topological transitions at $\omega=1$, which do not feature the EP. Instead, the change in knot topology is driven by a point of discontinuity in the spectrum. This discontinuity coincides precisely with the transition point observed in the Hermitian case. Spectral discontinuities for the systems defined by $A$ in Eq.~\eqref{HA-A-sigma-x} and \eqref{HB-A-sigma-x} of the main paper for the first and the second models respectively are demonstrated in the Fig.~\eqref{A11-HAHB-1}.

\section{Presence of an additional transition point}\label{appc}

\begin{figure}[htbp!]
  \begin{subfigure}{0.28\columnwidth}
  \includegraphics[width=\textwidth]{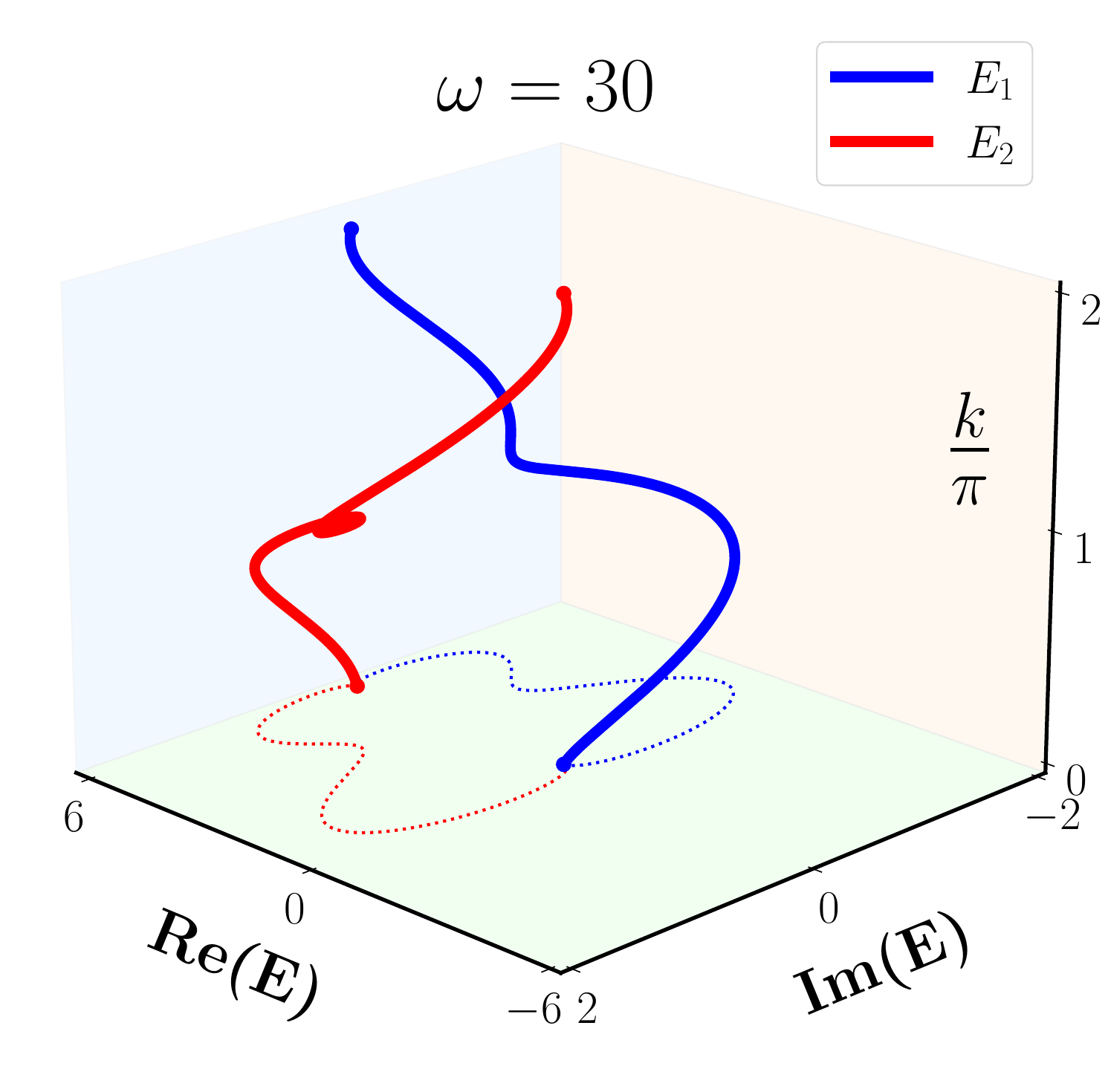}
  \caption{}\label{A11-30}
  \end{subfigure}
  \begin{subfigure}{0.28\columnwidth} 
  \includegraphics[width=\textwidth]{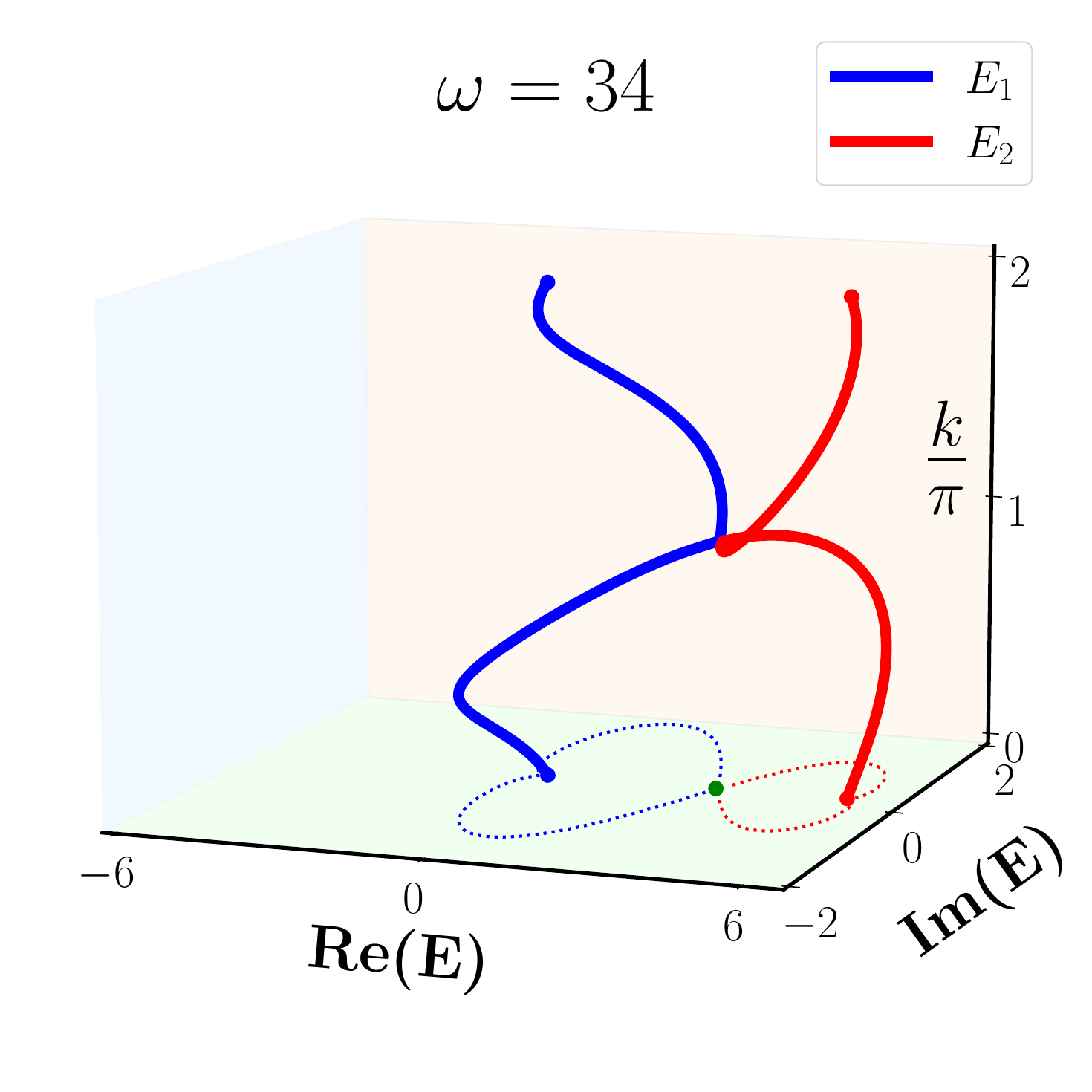} 
  \caption{}\label{A11-34} 
  \end{subfigure}
  \begin{subfigure}{0.28\columnwidth} 
  \includegraphics[width=\textwidth]{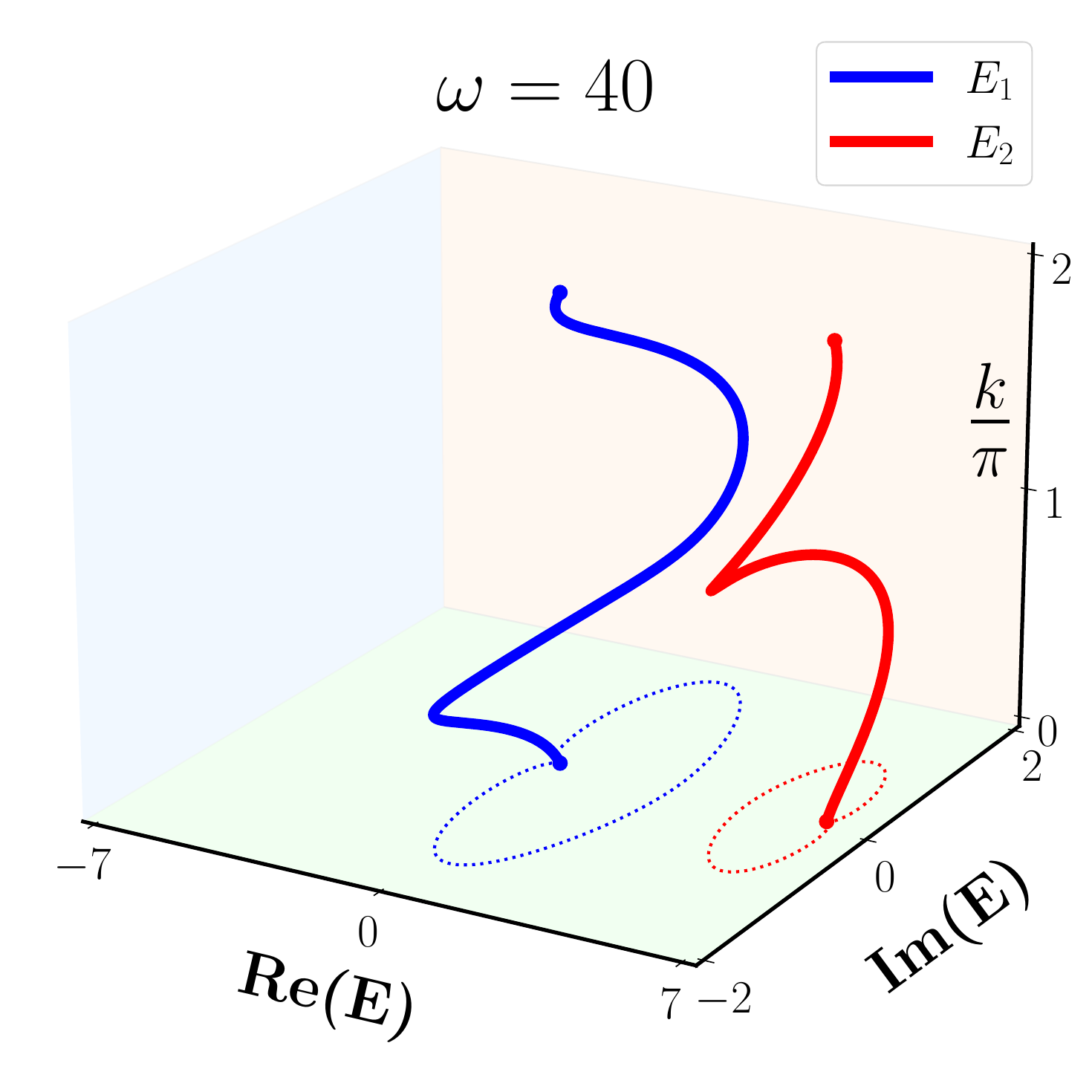} 
  \caption{}\label{A11-40} 
  \end{subfigure}
  \caption{
\emph{Knot structures of $A(\omega,k)$ for Model~I near the additional NH transition at $\omega \simeq 34$}. 
Panels (a) and (c) show the knot configurations for $\omega = 30$ and $\omega = 40$, respectively, 
while panel (b) corresponds to $\omega = 34$, where an exceptional point appears at $k = \pi$. 
This NH transition does not coincide with any Hermitian gap closing.
}\label{HA-34-new_ep}
\end{figure}

We present an important observation from the study of both models. As previously noted, topological transitions in the Hermitian case are directly related to those at the NH level, in the sense that both occur at the same value of the parameter, \( \omega = 1 \). Interestingly, in each of the two models with \( V = \sigma_x \), we found an additional point in the parameter space of \( \omega \): \( \omega\simeq34 \) for the first model and \( \omega\simeq 67 \) for the second model, where a topological transition takes place at the NH level. Consequently, we observe two distinct knot structures on either side of these two $\omega$ points in the respective models. Figure~\eqref{HA-34-new_ep} shows the two different knot structures in the regions \( 1 < \omega < 34 \) and \( \omega > 34 \) for model I, with $\omega=34$, $k=\pi$ representing the EP for this model.

In contrast to Fig.~\eqref{HA-34} of the main paper, which clearly shows that the configuration at \( \omega\simeq34 \) does not correspond to any topological transition in the Hermitian model $H^I$ (since there is no gap closing in the spectrum), Fig.~\eqref{HA-34-new_ep} highlights an important distinction in our analysis: while topological transitions at the Hermitian level are always accompanied by corresponding transitions at the NH level, the reverse is not necessarily true.

\begin{figure}[htbp!]
  \begin{subfigure}{0.28\columnwidth}
  \includegraphics[width=\textwidth]{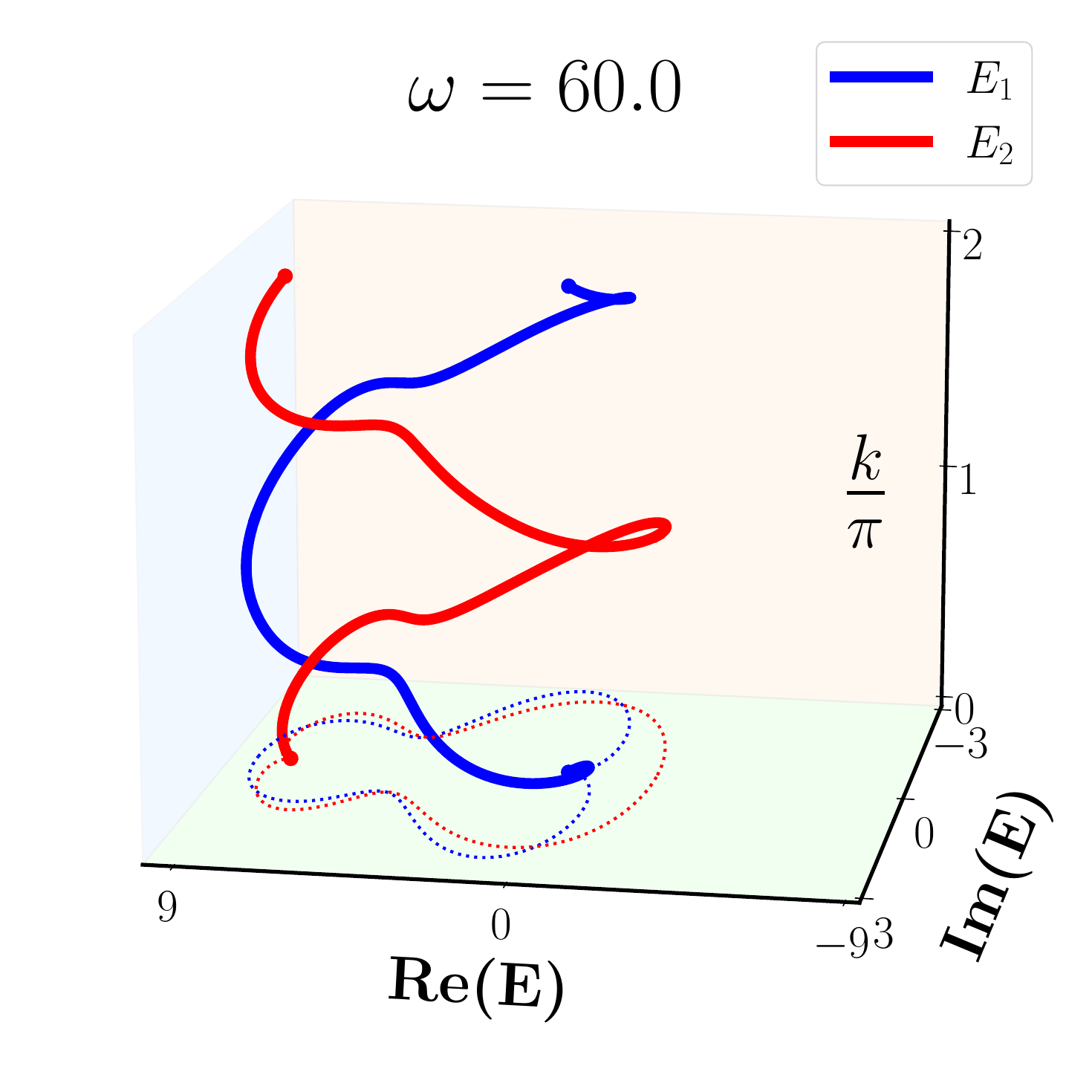}
  \caption{}\label{A11-60}
  \end{subfigure}
  \begin{subfigure}{0.28\columnwidth} 
  \includegraphics[width=\textwidth]{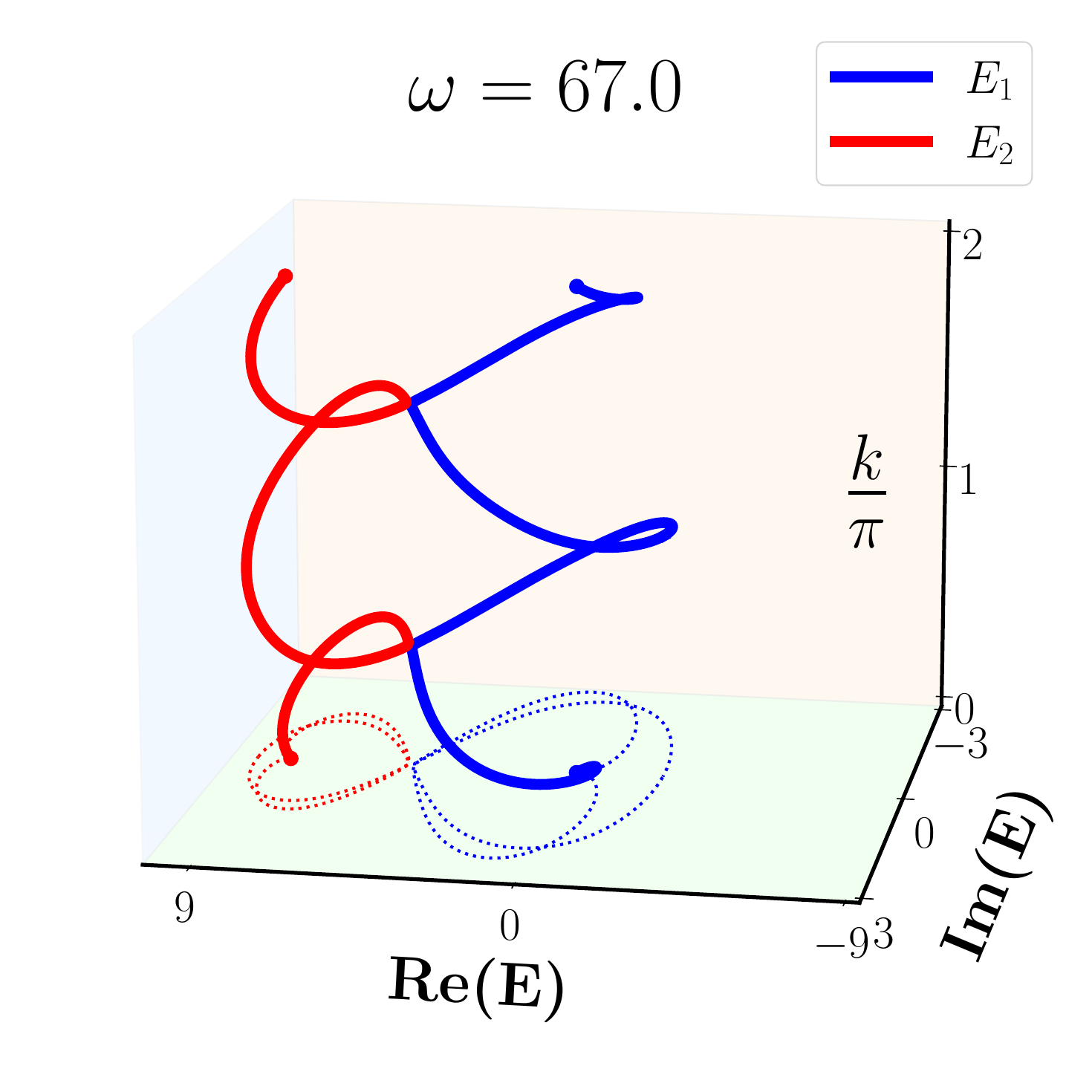} 
  \caption{}\label{A11-67} 
  \end{subfigure}
  \begin{subfigure}{0.28\columnwidth} 
  \includegraphics[width=\textwidth]{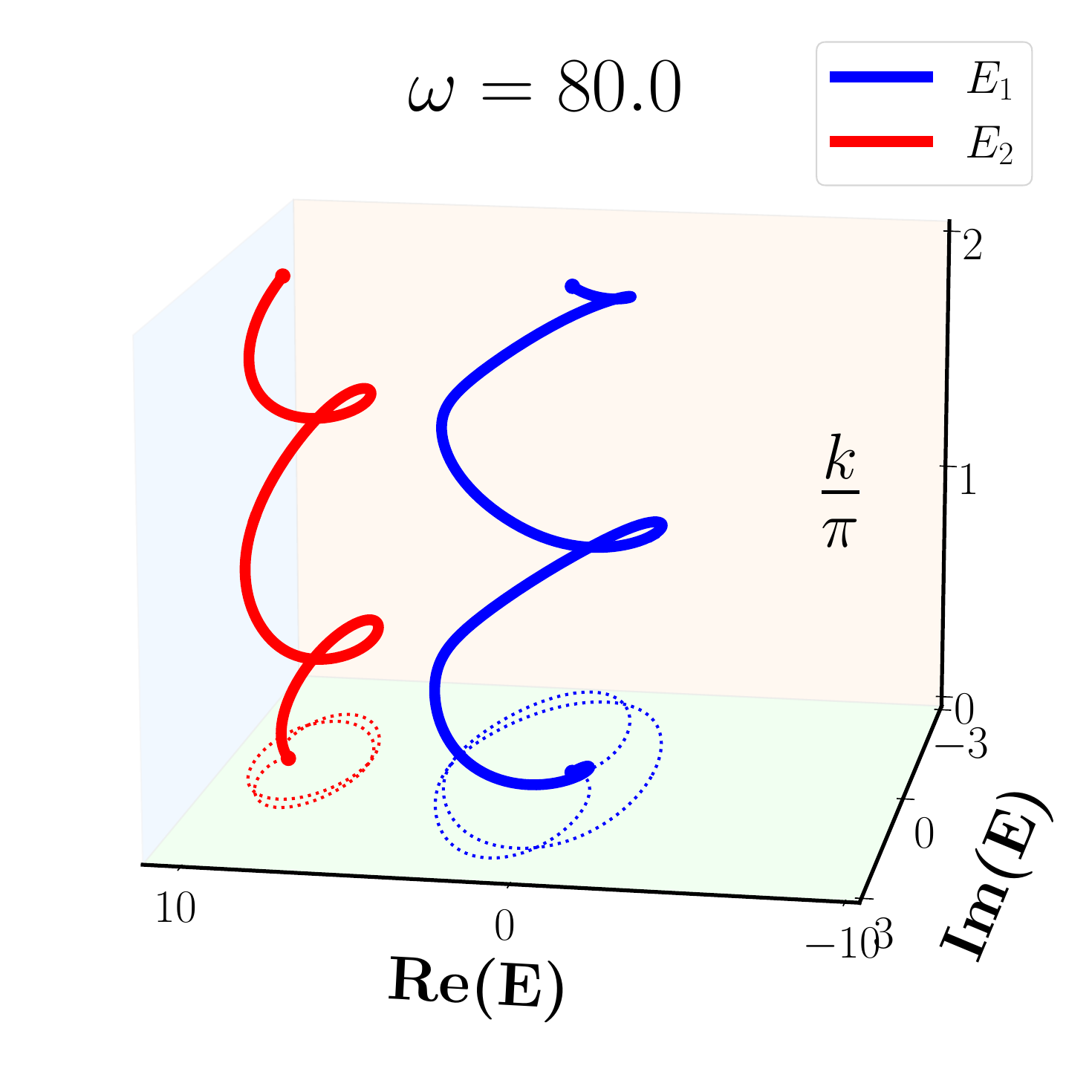} 
  \caption{}\label{A11-80} 
  \end{subfigure}
  \caption{\emph{Knot structures of $A(\omega,k)$ for Model~II near the additional NH transition at $\omega \simeq 67$}. 
Panels (a) and (c) show the configurations for $\omega = 60$ and $\omega = 80$, respectively, while panel (b) 
corresponds to $\omega = 67$, where a double exceptional point appears at 
$k = \frac{\pi}{2}$ and $k = \frac{3\pi}{2}$. 
This NH transition occurs without a corresponding Hermitian gap closing.
}\label{HB-67-new_ep}
\end{figure}

Similarly, Fig.~\eqref{HB-67-new_ep} shows two distinct knot structures in the regions \( 1 < \omega < 67 \) and \( \omega > 67 \) for model II, with $\omega=67$ corresponding to a double EP at $k=\frac{\pi}{2}$ and $\frac{3\pi}{2}$. Comparing this with Fig.~\eqref{HB-67}, we see that, for this model as well, the Hermitian counterpart exhibits no topological transition at $\omega=67$. This further confirms that while a Hermitian transition always implies a corresponding NH transition, the converse does not necessarily hold. We remark that exceptional points have recently been discussed in connection with van Hove singularities of Hermitian band structures~\cite{PhysRevB.111.L041405}. In the present SVD-based construction, although exceptional points appear in the spectrum of the NH matrices \(A(k)\), we do not find a corresponding band maximum or minimum in the spectrum of the parent Hermitian Hamiltonian. Consequently, the exceptional points observed here cannot be directly associated with van Hove singularities of the Hermitian system. Whether a more general relation between these two features exists remains an open question.


\section{First order knot transition for model II}\label{appc2}

\begin{figure}[htbp!]
    \centering \includegraphics[width=0.525\textwidth]{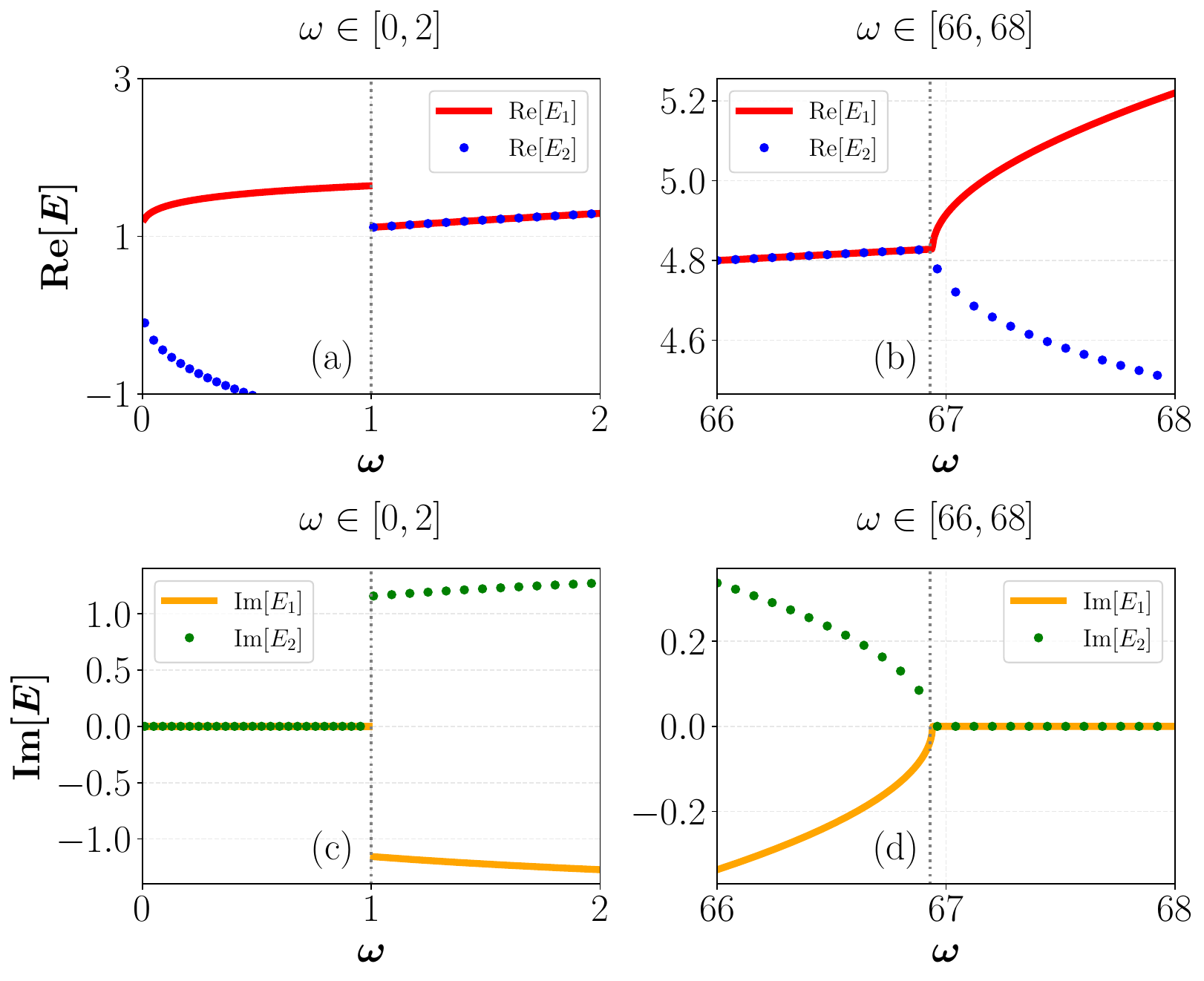}
    \caption{
\emph{Real and imaginary parts of the eigenvalues of $A(\omega,k)$ for Model~II}. Panels (a) and (c) show the discontinuous jump at $\omega = 1$, indicating the first-order knot transition. Panels (b) and (d) display the smooth behavior near the NH transition at $\omega \simeq 67$, where a double exceptional point occurs without any Hermitian gap closing.}\label{c-ec}
\end{figure}

For $V=\sigma_x$, we find in model II, similar to model I, a first-order knot transition at $\omega \simeq 67$, where both the real and imaginary parts show no discontinuity, as illustrated in Fig.~\eqref{c-ec}.

\section{Additional examples of $V$ for both models I and II}\label{appd}

We take up here another choice of $k$-independent traceless $V$ for each of the models and show that even in this case the nature of the topological transition still remains the same as for the earlier choices, i.e., with the new choice of $V$ we still get an unlink-unknot transition for the model I and an unknot-hopflink transition for model II.
  
\begin{figure}[htbp!]
  \begin{subfigure}{0.3\columnwidth}
  \includegraphics[width=\textwidth]{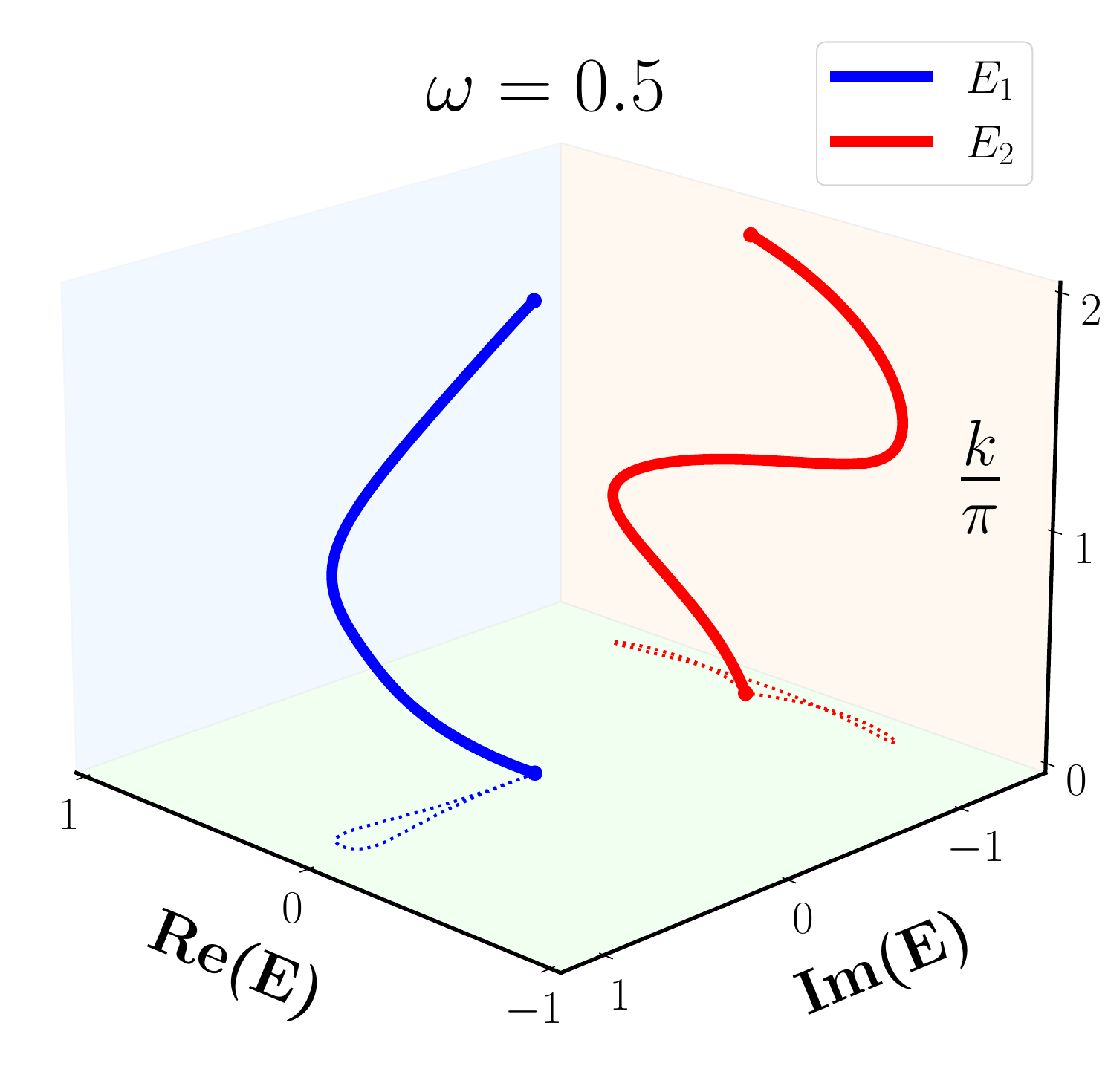}
  \caption{}\label{A55-0_5-unlink}
  \end{subfigure}
  \begin{subfigure}{0.3\columnwidth} 
  \includegraphics[width=\textwidth]{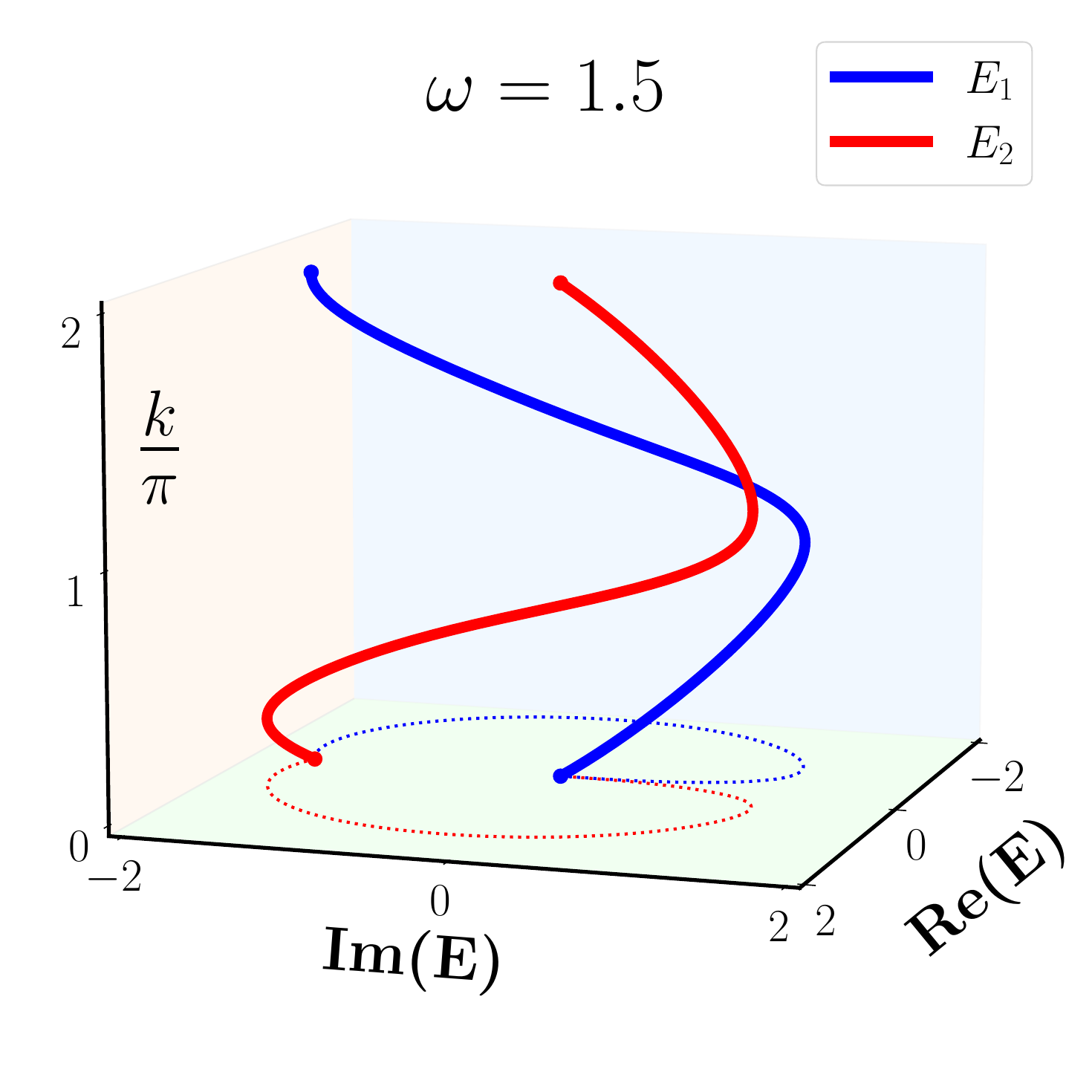} 
  \caption{}\label{A55-1_5-unknot} 
  \end{subfigure}
  \caption{
\emph{Knot structures of $A(\omega,k)$ for Model~I with the traceless, $k$-independent choice $V = i\sigma_z$}. Panel (a) shows the unlink for $\omega = 0.5$, while panel (b) shows the unknot for $\omega = 1.5$. This demonstrates that the unlink–unknot transition persists for this alternative choice of $V$.}\label{A55-HA}
\end{figure}

To demonstrate this, we consider $V=i\sigma_z$ for both the models, where $\sigma_z$ is a Pauli matrix. The NH system $A$ for model I is then given by
\begin{equation}\label{HA-A-V1}
    A=\begin{pmatrix}
-\frac{i e^{i k} a \sqrt{1 + \omega + a}}{\sqrt{2} (e^{i k} + \omega)} &
-\frac{i e^{i k} a \sqrt{1 + \omega - a}}{\sqrt{2} (e^{i k} + \omega)} \\
-\frac{i \sqrt{1 + \omega + a}}{\sqrt{2}} &
\frac{i \sqrt{1 + \omega - a}}{\sqrt{2}}
\end{pmatrix},
\end{equation}
where, $a=\sqrt{1 + \omega^2 + 2 \omega \cos k}$. The knot structures for this case are shown in Fig.~\eqref{A55-HA}.

\begin{figure}[htbp!]
  \begin{subfigure}{0.3\columnwidth}  \includegraphics[width=\textwidth]{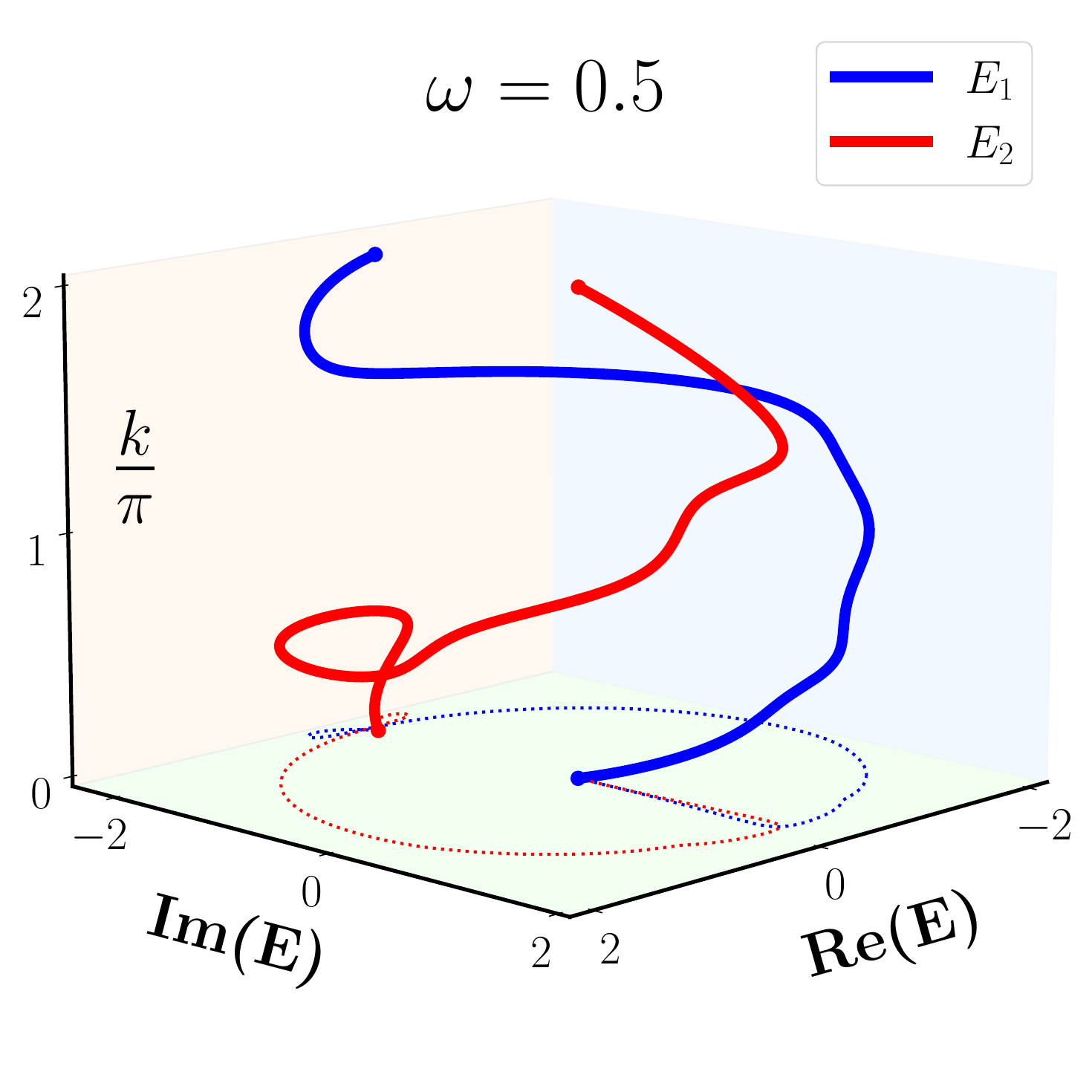}
  \caption{}\label{A55-0_5-unknot}
  \end{subfigure}
  \begin{subfigure}{0.3\columnwidth} 
\includegraphics[width=\textwidth]{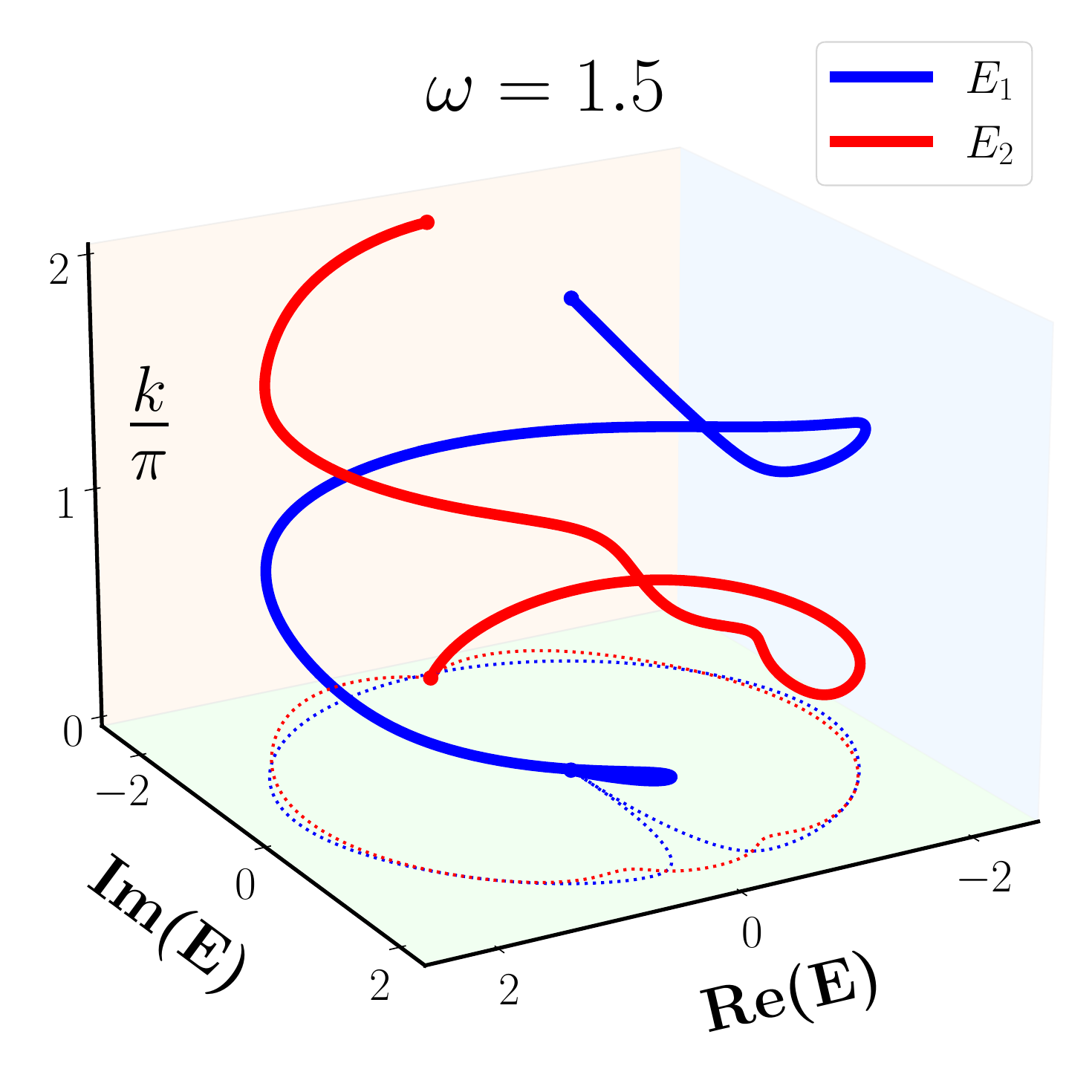} 
  \caption{}\label{A55-1_5-hopflink} 
  \end{subfigure}
  \caption{
\emph{Knot structures of $A(\omega,k)$ for Model~II with the traceless, $k$-independent choice $V = i\sigma_z$}. Panel (a) shows the unknot for $\omega = 0.5$, while panel (b) shows the Hopf-link for $\omega = 1.5$. This confirms that the unknot–hopflink transition of Model~II is preserved for this choice of $V$.}\label{A5-HB}
\end{figure}

While, for model II, the NH system $A$  with the choice, $V=i\sigma_z$ is given by,

\begin{equation}
    A=\begin{pmatrix}
-\frac{i a' \sqrt{
\left(3 + \omega + a'\right)}}{\sqrt{2} (1 + e^{2i k} \omega + 2 \cos k)}
& 
-\frac{i a'\sqrt{ 
\left(3 + \omega - a'\right)}}{\sqrt{2} (1 + e^{2i k} \omega + 2 \cos k)} \\[10pt]
-\frac{i \sqrt{3 + \omega + a'}}{\sqrt{2}} 
& 
\frac{i \sqrt{3 + \omega - a'}}{\sqrt{2}}
\end{pmatrix},
\end{equation} 
where $a'=\sqrt{3 + \omega^2 + 2\left(\left(2+\omega\right)\cos k + \left(1 + \omega\right) \cos 2k + \omega \cos 3k\right)}$. The knot structures for this system for $\omega<1$ and $\omega>1$ are demonstrated in Fig.~\eqref{A5-HB}.

\newpage

\begin{figure}[htbp!]
  \begin{subfigure}{0.3\columnwidth}
  \includegraphics[width=\textwidth]{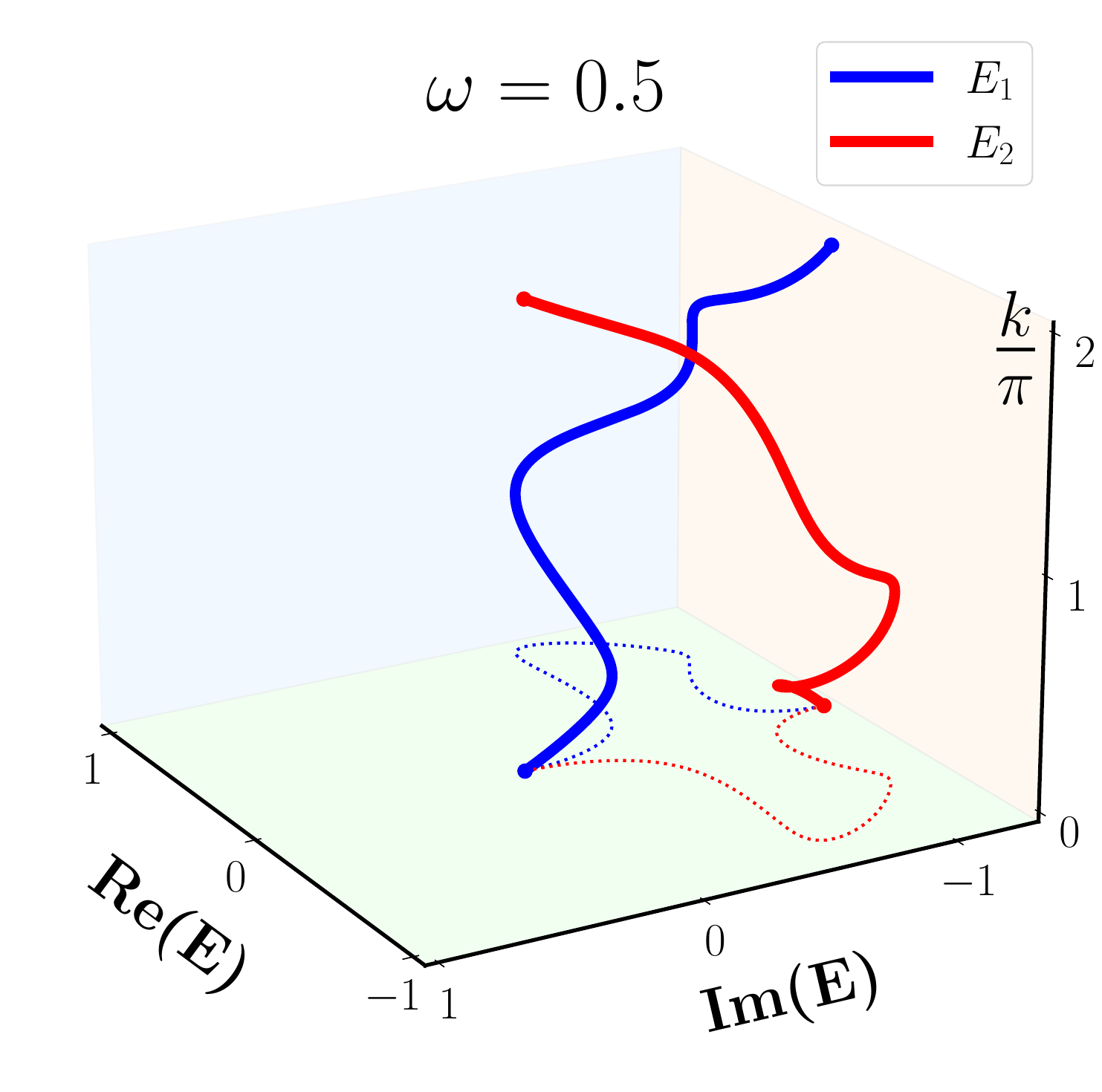}
  \caption{}\label{A22-0_5-unknot}
  \end{subfigure}
  \begin{subfigure}{0.3\columnwidth} 
  \includegraphics[width=\textwidth]{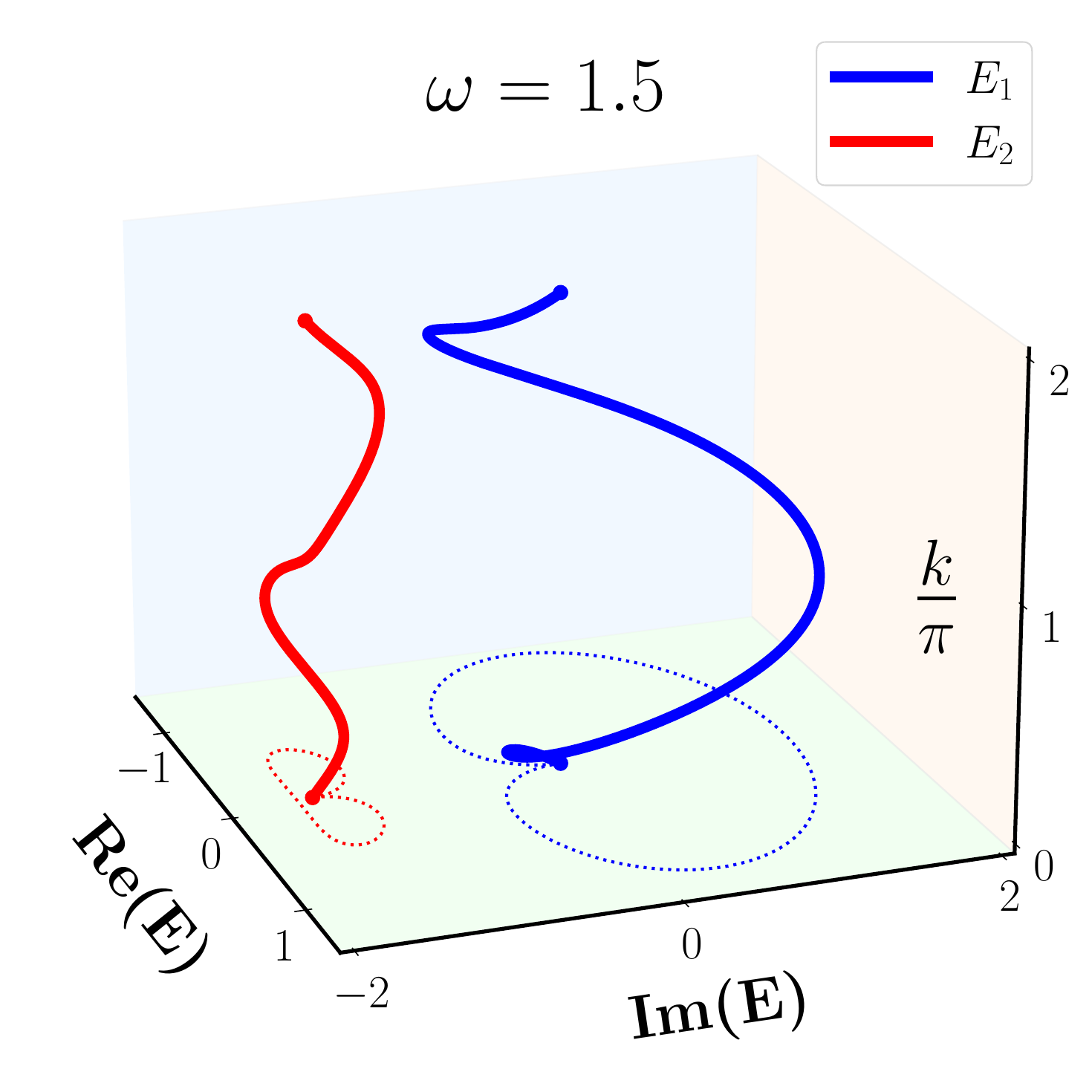} 
  \caption{}\label{A22-1_5-unlink} 
  \end{subfigure}
  \caption{\emph{Knot structures for Model~I with $V=\sigma_y$}. Panel (a) shows an unknot at $\omega=0.5$, while panel (b) shows an unlink at $\omega=1.5$. This choice of $V$ exhibits a reversed knot transition as $\omega$ crosses the Hermitian topological transition point.}\label{A22-HA}
  \end{figure}

However, apart from this, there exists certain choices of $k$-independent traceless $V$, for which the knot transition can occur in the reverse direction. To demonstrate this, we revisit the first model. As already observed, for $V = \sigma_x$ and $i\sigma_z$, the transition was from unlink to unknot for \( \omega < 1 \) and \( \omega > 1 \), respectively, however, in contrast, for \( V = \sigma_y \), the transition occurs from unknot to unlink in the regions \( \omega < 1 \) and \( \omega > 1 \), as shown in Fig.~\eqref{A22-HA}.
For this case, the matrix \( A \), computed using Eq.~\eqref{SVD} of the main paper, takes the form:
\begin{equation}
    A=\begin{pmatrix}
-\frac{i e^{i k} a\sqrt{1 + \omega - a}}{\sqrt{2} (e^{i k} + \omega)} & -\frac{i e^{i k} a \sqrt{1 + \omega + a}}{\sqrt{2} (e^{i k} + \omega)} \\
\frac{i \sqrt{1 + \omega - a}}{\sqrt{2}} & -\frac{i \sqrt{1 + \omega + a}}{\sqrt{2}},
\end{pmatrix}.
\end{equation}
where a=$\sqrt{1 + \omega^2 + 2 \omega \cos{k}}$

 
\section{Model I: Effective real-space NH description of $A$ near knot transition}\label{appdF}
  In this section, we demonstrate the effective real-space NH description of the $A$ matrix for Model~I near knot transition point. The explicit form of $A(\omega,k)$ is given in 
Eq.~\eqref{SVD}. While the real-space version of this matrix can appear extremely complicated, we expand it at  $\omega=1$, in the neighborhood of $k=\pi$, where the knot transition occurs. 
In the linear order of $k$, $A(k,\omega=1)$ reads as, 
\begin{equation}\label{A-+final}
A(k=\pi+,\omega=1) \approx 
\begin{pmatrix}
(k-\pi)+i\left(1 - \tfrac{k - \pi}{4}\right) & (k-\pi)-i\left(1 + \tfrac{k - \pi}{4}\right) \\[6pt]
1 - \tfrac{k - \pi}{4} & 1 + \tfrac{k - \pi}{4}
\end{pmatrix}.
\end{equation}
and,
\begin{equation}\label{A-final-minus}
A(k=\pi-,\omega=1) \approx 
\begin{pmatrix}
-\,i + (k-\pi)\left(1 - \tfrac{i}{4}\right) &
\; i + (k-\pi)\left(-1 - \tfrac{i}{4}\right) \\[6pt]
1 + \tfrac{k - \pi}{4} & 1 - \tfrac{k - \pi}{4}
\end{pmatrix}.
\end{equation}
The same (up to linear order in $k$) $A$ matrix can be obtained by doing a linear expansion of $k$ around $k=\pi$, where $A_{\text{eff}}=M\sin k+C$, 
where
\[
C(k=\pi+) =
\begin{pmatrix}
i & -i \\[4pt]
1 & 1
\end{pmatrix},
\qquad
M(k=\pi+) =
\begin{pmatrix}
-1 + \tfrac{i}{4} & -1 + \tfrac{i}{4} \\[6pt]
\tfrac14 & -\tfrac14
\end{pmatrix},
\]
and
\[
C(k=\pi-) =
\begin{pmatrix}
-\,i & i \\[4pt]
1 & 1
\end{pmatrix},
\qquad
M(k=\pi-) =
\begin{pmatrix}
-1 + \tfrac{i}{4} & 1 + \tfrac{i}{4} \\[6pt]
-\tfrac14 & \tfrac14
\end{pmatrix}.
\]
Now, it is reasonably straightforward to perform a Fourier transformation of $A_{\text{eff}}$. The real-space representation of $A_{\text{eff}}$ then reads as,
\begin{equation}
A_{\text{eff}} = \sum_n \left[ \frac{1}{2i} \left( c_{n+1}^\dagger M c_n - c_n^\dagger M c_{n+1} \right) + c_n^\dagger C c_n \right].
\end{equation}
Here, $c_n$ is the two-component vector of sublattice operators in the $n$-th unit cell, $M$ is the $2\times 2$ hopping matrix, and $C$ is the $2\times 2$ on-site matrix. It is a NH lattice Hamiltonian with short-range, nearest-neighbor hopping, featuring complex hopping amplitudes and complex on-site potentials, having two sub-lattices.
Matrix $A$ and $C$  have finite discontinuity at $k=\pi$. This remains the underlying source of the \textit{first-order knot transition} reported in our manuscript. Although $A$ and $A_{\text{eff}}$ are two distinct NH models, their expansions in the vicinity of the knot–transition point are very similar. Consequently, both $A$ and $A_{\text{eff}}$ are expected to exhibit similar physics around $\omega = 1$ and $k = \pi$.

\end{document}